\documentclass[aps,prb,showpacs,preprintnumbers,nofootinbib,twocolumn,superscriptaddress]{revtex4-2}
\usepackage{amsmath,amssymb}
\usepackage{graphicx}
\usepackage{enumitem} 
\usepackage{array}
\usepackage{physics}
\usepackage[colorlinks,bookmarks=true,citecolor=blue,linkcolor=red,urlcolor=blue]{hyperref}

\newcommand{\pdag}{^{\phantom{\dagger}}}
\renewcommand{\dag}{^{\dagger}}
\newcommand{\proj}[1]{P_{\langle #1\rangle}}
\newcolumntype{P}[1]{>{\centering\arraybackslash}p{#1}}
\newcommand{\ceqn}[1]{Eq.~(\ref{eq:#1})}
\newcommand{\cfig}[1]{Fig.~\ref{fig:#1}}

\begin{document}
	\title{Supersymmetry on the honeycomb lattice:\\ resonating charge stripes, superfrustration, and domain walls}
	\author{Patrick H. Wilhelm}	
	\thanks{These two authors contributed equally}
	\affiliation{Institut für Theoretische Physik, Universität Innsbruck, A-6020 Innsbruck, Austria}
	\author{Yves H. Kwan}
	\thanks{These two authors contributed equally}
	\affiliation{Princeton Center for Theoretical Science, Princeton University, Princeton NJ 08544, USA}
	\affiliation{Rudolf Peierls Centre for Theoretical Physics, Parks Road, Oxford, OX1 3PU, UK}
	\author{Andreas M. L\"{a}uchli}
	\affiliation{Laboratory for Theoretical and Computational Physics, Paul Scherrer Institute, 5232 Villigen, Switzerland}
	\affiliation{Institute of Physics, \'{E}cole Polytechnique F\'{e}d\'{e}rale de Lausanne (EPFL), 1015 Lausanne, Switzerland}
	\author{S.A. Parameswaran}
	\affiliation{Rudolf Peierls Centre for Theoretical Physics, Parks Road, Oxford, OX1 3PU, UK}
	
	\begin{abstract}
		We study  a model of spinless fermions on the honeycomb lattice with nearest-neighbor exclusion and extended repulsive interactions that exhibits `lattice supersymmetry' [P. Fendley, K. Schoutens, and J. de Boer, {\it Phys. Rev. Lett.} {\bf 90}, 120402 (2003)]. Using a combination of exact diagonalization of large ($N\leq56$ site) systems, mean-field numerics, and symmetry analysis, we establish a rich phase structure as a function of fermion density, that includes non-Fermi liquid behavior, resonating charge stripes, domain-wall and bubble physics, and identify a finite range of fillings with extensive ground state degeneracy and both gapped and gapless spectra. We comment on the stability of our results to relaxing the stringent requirements for supersymmetry, and on their possible broader relevance to systems of strongly-correlated electrons with extended repulsive interactions.
	\end{abstract}
	
	\maketitle

	\section{Introduction}
	
	Quantum models with constrained Hilbert spaces  often arise as special  limits of strongly correlated many-body systems. A classic example is the $t$-$J$ model, which has been invoked to study the interplay of superconductivity and antiferromagnetism in the high-$T_c$ cuprate materials. Here, the low-energy spectrum is characterized by a  prohibition on doubly-occupied sites   that arises naturally in the large-$U$ limit of the spinful Hubbard model. As a result of this constraint, the only low-energy degrees of freedom are holes (unoccupied sites) and the spins of singly-occupied sites. Models with non-onsite constraints can also be physically relevant: a case in point is the so-called `PXP' model of a chain of atoms which can be either be in their ground or excited states, with the constraint that nearest-neighbour atoms cannot be simultaneously excited. An effective Hamiltonian with this `blockade' constraint emerges in arrays of Rydberg atoms due to the strong short-distance van der Waals repulsion between Rydberg-excited states. Twisted bilayer graphene~\cite{Cao2018correlated,Cao2018unconventional,Yankowitz2019tuning,Lu2019orbital} motivates another recent example: here, the topology and symmetries of the flat bands require that their Wannier orbitals form `fidget spinners' lying on the sites of a honeycomb lattice, but with charge delocalized on the centers of the three adjacent hexagonal~\cite{Koshino2018maximally,Kang2018symmetry,Zou2018band} plaquettes. For sufficiently strong local repulsion, the effect of the resulting cluster charging energy can be approximated by an effective
	hard-core exclusion that extends to third neighbour sites~\cite{Zhang2022fractional,Mao2022fractionalization}.
	\begin{figure}
		\centering
		\includegraphics[width=1\columnwidth]{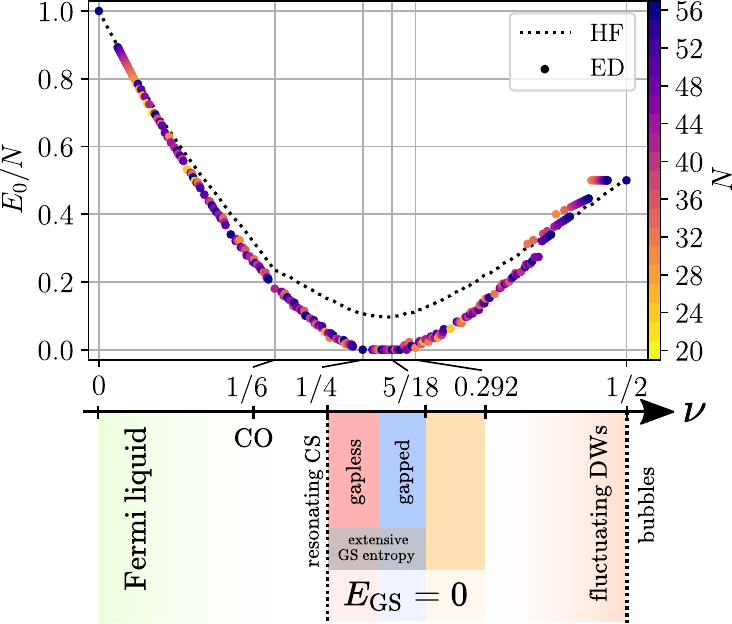}
		\caption{\textbf{Ground state energy density $E_0/N$ vs filling $\nu$, and schematic phase diagram for the SUSY honeycomb model.} Exact diagonalization (ED) results are shown for varying number of sites $N$, while Hartree-Fock (HF) energies are given for a system with $12\times12$ unit cells. Zero-energy states are found within the filling window $1/4\leq\nu\lesssim 0.3$, which is further subdivided into different regimes. CO: charge order, CS: charge stripes, DW: domain wall.}
		\label{fig:phase_diagram_schematic}
	\end{figure}
	
	An apparently different class of  constrained interacting models, where  `hard-core' fermions obey nearest-neighbour exclusion, has been introduced and explored by Fendley and Schoutens with various  collaborators~\cite{Fendley2003lattice,Fendley2003bethe,fendley2005susy,huijse2008review} (FSX). A peculiar feature of the FSX Hamiltonians is that they obey a supersymmetric (SUSY) quantum mechanics~\cite{Witten1982constraints}. This additional symmetry imposes additional structure on the many-body spectrum, for instance, allowing the derivation of non-trivial lower bounds on the ground state degeneracy (GSD). For several geometries, such as the triangular and honeycomb lattices, the GSD has been predicted to be exponential in the total area~\cite{vaneerten2005witten}, and the existence of zero-energy states has been proven for a range of fillings~\cite{jonsson2010homology}. 
	Such models are promising for obtaining unusual correlated phenomena, since the extensive nature of the ground state entropy, dubbed \emph{superfrustration} \cite{fendley2005susy,huijse2008susy,huijse2008review,Huijse2012triangular,Huijse2011staggered}, over a finite filling window suggests a multi-criticality between several competing orders~\cite{Huijse2012triangular,Bauer2013multicritical,chepiga2021ladder}. A prior numerical study has confirmed the exponential degeneracy on the triangular lattice, and reported unusual behaviors within the regime of zero-energy states~\cite{galanakis2012triangular}. 
	
	Although the first class of problems seems more `natural' as they do not require the special structure inherent in SUSY models, the distinction is not necessarily  so clear-cut. For instance, special limits of the $t$-$J$ model enjoy a symmetry superalgebra~\cite{Sarkar1991tJ,Essler1992tJ,Sarkar1991supercoherent}, while the PXP model of Rydberg blockade is closely related to the one-dimensional FSX model. Less directly, constraining single-particle hopping terms to act within the manifold of states that satisfy the constraint of no-double-occupancy of a fidget-spinner Wannier orbital leads to effective Hamiltonians reminiscent of the honeycomb lattice FSX model. Given the challenges inherent to studying strongly-correlated many-particle systems, proximity to a SUSY model could provide a new starting point from which to access the physics of more realistic parameter regimes, where (approximate) SUSY could serve as an organizing principle to understand the competition between different phases.
	
	With this broad motiviation, here we focus on the supersymmetric FSX model on the honeycomb lattice. This is the last of the simple two-dimensional lattices, yet has eluded a thorough characterization to date. 
	This is partly because, compared to the square and triangular cases, the lower coordination number of the honeycomb lattice --- which controls the severity of the hard-core constraint --- combined with the added sublattice degree of freedom makes a numerical treatment more challenging. However, the inclusion of sublattice structure can also give rise to new phenomena, as  demonstrated by the bubble and domain wall physics that we explore below (Sec.~\ref{sec:highfilling}).
	Our investigation of the model (described in Sec.~\ref{subsec:model}) is rooted in a comprehensive exact diagonalization (ED) study resolving all spatial symmetries, and supplemented with analytical calculations and Hartree-Fock numerics, as outlined in Sec.~\ref{subsec:methods}. 
	
	We obtain a rich phase diagram with several distinct filling regimes, including a zero-energy filling\footnote{We define the filling $\nu$ to be the number of fermions per {\it site} rather than per unit cell; while the latter is the more correct choice and is relevant e.g. to band theory, the former is more convenient for discussing real-space charge orders.} window, as summarized in Fig.~\ref{fig:phase_diagram_schematic} and Fig.~\ref{fig:zero_degeneracy_spectra}. Strikingly, the phase structure exhibits phenomenology characteristic of both the previously-studied square and triangular lattice FSX models. We now summarize our key results, which also serves to lay out the organization of this paper, before closing this introduction by defining our model and outlining our numerical methods.
	
	In Sec.~\ref{sec:lowfilling}, we first discuss the more conventional phases at low fillings, which consist primarily of a Fermi liquid interspersed with charge order (CO) at commensurate densities. 
	In Sec.~\ref{sec:zeroenergy}, we study the zero-energy window at intermediate fillings in detail. We uncover a resonating charge-stripe ordered phase at filling $\nu=1/4$. This phase is compressible under the introduction of line-sublattice flips and hence exponentially degenerate in the linear dimension of the system, thereby reproducing features reminiscent of the SUSY model on the square lattice~\cite{huijse2008susy}. Furthermore, our results confirm superfrustration in the filling window $0.25 < \nu \lesssim 5/18$, where the extensivity of the ground state entropy as well as the presence of a spectral gap are found to be filling-dependent, as in the triangular lattice~\cite{Huijse2012triangular,galanakis2012triangular}. Interestingly, we find an unbroken zero-energy window for all rational fillings considered up to $\nu\simeq 0.286$, and robust zero-energy states for fillings as high as $\nu=7/24\simeq 0.292$--- significantly surpassing the conjectured upper bound  $\nu=5/18 \simeq 0.278$ for this window, computed in Ref.~\cite{jonsson2010homology} using homology. 
	In Sec.~\ref{sec:highfilling}, we investigate the physics at high fillings, which is characterized by bubbles and domain walls of sublattice polarization, and discuss its relation to the dense liquid immediately above the zero-energy window. We close with a  discussion of our results and possible future directions  in Sec.~\ref{sec:conclusions}.
	
	\subsection{Model and Known Results}\label{subsec:model}
	
	Supersymmetric quantum mechanics is described by a positive semi-definite Hamiltonian and nilpotent supercharge operators which satisfy the following algebra
	\begin{equation}
		\label{eq:H_susy_general}
		\begin{gathered}
			H=\{Q,Q^\dagger\}\\
			[H,Q]=[H,Q^\dagger]=0\\
			\{Q,Q\}=\{Q^\dagger,Q^\dagger\}=0.
		\end{gathered}
	\end{equation}
	Consider spinless fermions on some lattice with sites $i$. Following Fendley and Schoutens~\cite{fendley2005susy,huijse2008review}, we can generate a number-conserving interacting fermion model by choosing $Q^\dagger=\sum_i c^\dagger_i P_{\langle i\rangle}$, where the projector $P_{\langle i \rangle}=\prod_{j\text{ next to }i}(1-n_j)$. [This is sometimes referred to as the `$M_1$ model~\cite{Fendley2003bethe}'; we will refer to this as the SUSY model below.] We work in the Hilbert subspace without nearest neighbour occupation, such that the dynamics is intuitively cast in terms of extended hard-core fermions.
	
	\begin{figure}[htp]
		\centering
		\includegraphics[width=.3\textwidth]{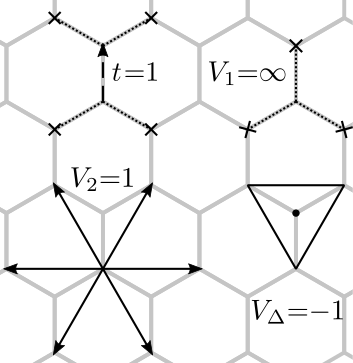}
		\caption{\textbf{Hopping and interaction terms.} Values corresponding to the SUSY limit are indicated.}
		\label{fig:interactions}
	\end{figure}
	
	For the honeycomb lattice with $N$ sites and $N_e$ fermions, the Hamiltonian then reads (see Fig.~\ref{fig:interactions})
	\begin{align}
		\label{eq:H_susy_honeycomb}
		H &= \sum_{\langle i j \rangle} \proj{i} c_i\dag c_j\pdag \proj{j} + \sum_i \proj{i} \nonumber \\
		&= \sum_{\langle i, j \rangle} \proj{i}^a a_{i}\dag b_{j}\pdag \proj{j}^b + \mathrm{H.c.} + \sum_{\alpha} \sum_{\langle \langle i,j \rangle \rangle} n_i^{\alpha} n_j^{\alpha} \nonumber \\
		& \quad - \sum_{ i,j,k \in \Delta} n_i^{\alpha} n_j^{\alpha} n_k^{\alpha} + N - 3 N_e
	\end{align}
	where $a^\dagger$ ($b^\dagger$) is the creation operator for sublattice $\alpha = A$ ($B$). Up to an overall shift $N$, the model consists of nearest neighbour (n.n.) hopping $t=1$ of hard-core fermions, n.n.n. repulsion $V_2=1$, an attractive three-body attraction $V_\triangle=-1$, and a chemical potential term $\mu=-3$. Besides $U(1)$ charge conservation, the Hamiltonian of \ceqn{H_susy_honeycomb} is translationally invariant and respects  $D_6$ point group of the honeycomb lattice. Note that the classical interaction energy (obtained by setting $t=0$ in \ceqn{H_susy_honeycomb}) of any Fock state is positive. The hard-core condition constrains the filling factor to lie in $\nu\in [0,\frac{1}{2}]$. 
	Using the SUSY algebra, it can be shown that positive energy states come in doublets that differ in fermion number $N_e$ by 1, while zero-energy ground states appear as singlets. Therefore the Witten index $W=\text{tr}(-1)^{N_e}$ simply counts the difference between the number of ground states with odd and even fermion parity, and provides a lower bound for the total GSD~\cite{Witten1982constraints}. Using transfer matrix methods, Van Eerten~\cite{vaneerten2005witten} has estimated an exponential in system size scaling of $|W|\propto (1.2\pm 0.1)^N$, implying an extensive ground state entropy (GSE). Furthermore, Jonsson~\cite{jonsson2010homology} has used homology to demonstrate the existence of zero-energy states at all fillings $\nu\in [\frac{1}{4},\frac{5}{18}]$ in the thermodynamic limit (TDL), though this does not rule out ground states outside this filling range.  
	
	\subsection{Numerical Methods}\label{subsec:methods}
	
	\begin{figure*}[htp]
		\includegraphics[width=\textwidth]{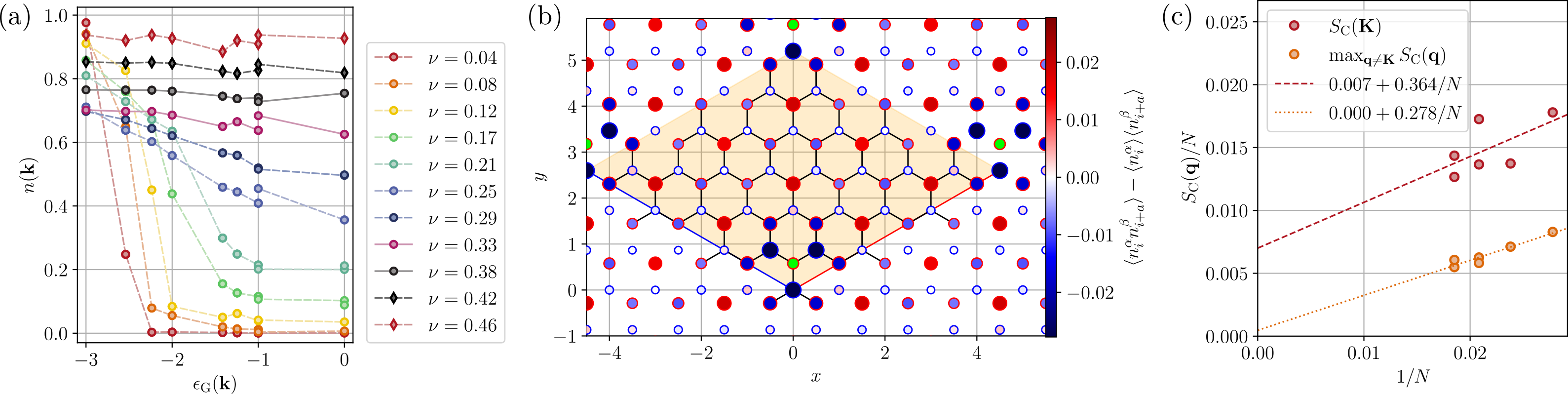}
		\caption{\textbf{Fermi liquids and charge order at low filling.} (a) The momentum-space occupation number $n(\mathbf{k})$ from ED as a function of the graphene dispersion $\epsilon_{\mathrm{G}}(\mathbf{k})$ exhibits a (softened) step at fillings $\nu < 1/4$, indicative of a Fermi liquid. (b) At $\nu=1/6$, the connected density-density correlation function relative to the site marked in green shows clear signatures of sublattice polarized CO with a tripled unit cell. (c) The long-range character is substantiated by the finite value of the extrapolation of the static charge structure factor peak at $\mathbf{q}=\mathbf{K}$ to the TDL. For fillings with solid lines in (a), the ground state is completely unique, while for dashed ones it is only unique in its symmetry sector. Red/blue outlined circles in (b) correspond to sites on the  $A$/$B$ sublattice. The clusters used are $\mathbf{u}_1=(1,4),\mathbf{u}_2=(5,-4),N=48$ for (a), and $\mathbf{u}_1=(3,3),\mathbf{u}_2=(3,-6),N=54$ for (b).}
		\label{fig:low_filling}
	\end{figure*}
	
	In the search for ground states of $H$, the form of \ceqn{H_susy_general} suggests to look into the properties of the (non-hermitian) supercharge operator $Q$. Indeed, for any ground state $\ket{\Psi}$ of the SUSY Hamiltonian, it must hold that it is in the kernel of $Q$ ($Q \ket{\Psi}=0$), but not in its image. Hence, one could in principle obtain the number of zero-energy modes as the dimension of this subspace $\mathcal{H}_Q = \mathrm{ker}\,Q/\mathrm{Im}\,Q$, i.e. the cohomology of $Q$ \cite{Fendley2003SUSY,fendley2005susy,huijse2008susy}. Practically, however, on a generic lattice it is difficult to find $\mathcal{H}_Q$ analytically, while numerical methods greatly benefit from the  hermicity  of $H$, absent in $Q$.
	For this reason, our primary method for the investigation of the ground state properties of $H$ is real-space ED on finite systems with periodic boundary conditions. We combine a thorough analysis of the low-energy many-body spectrum, which carries crucial information about e.g. spontaneous symmetry breaking, with direct measurements of observables in the many-body ground state. Crucially, we exploit both $U(1)$ charge conservation and all lattice symmetries (translations and rotations) as well as an exact implementation of the nearest-neighbor hard-core constraint. This enables us to reach system sizes up to $N=56$ throughout the whole filling range. Furthermore, the decomposition of the full Hilbert space into subspaces corresponding to irreducible representations of the underlying symmetry group is at the heart of the symmetry analysis of the low-energy spectrum described in Ref.~\cite{wietek2017tos}, which we have adapted to fermions for this study. Together with the large (symmetry-resolved) Hilbert space dimensions $\sim 10^8-10^9$, degeneracies in the zero-energy window (within one symmetry sector) limit the ARPACK-based diagonalization \cite{Lehoucq1998arpack} due to increased memory requirements. In order to keep our setup as unbiased as possible, unless mentioned otherwise, we work with all possible symmetry-unrelated simulation clusters of a given number of sites $N$ with an aspect ratio (i.e. the length of the longest over the shortest straight loop around the torus) no greater than 1.5. The vectors of the simulation cell are given by $\mathbf{L}_1 = \mathbf{u}_1 \cdot \bar{\mathbf{a}}$, $\mathbf{L}_2 = \mathbf{u}_2 \cdot \bar{\mathbf{a}}$, where $\bar{\mathbf{a}} \equiv (\mathbf{a}_1, \mathbf{a}_2)^{\mathrm{T}}$,  $\mathbf{u}_i \in \mathbb{Z}^2$ and $\mathbf{a}_1=(1,0),\mathbf{a}_2=(\frac{1}{2},\frac{\sqrt{3}}{2})$ are the honeycomb lattice primitive vectors. 
	
	To supplement the ED results, we also perform a self-consistent Hartree-Fock (HF) study of the ground state at all fillings. This is useful for diagnosing charge ordering tendencies, as well as estimating the importance of beyond mean-field corrections, by comparing with ED. We restrict to periodic boundary conditions and $\mathbf{L}_i=L_i \mathbf{a}_i$, again with a maximum aspect ratio of 1.5. The hard-core constraint is implemented by adding a large n.n. repulsion $V_1\rightarrow\infty$. Given the translation symmetry, a natural choice is to consider one-body density matrices that are diagonal in momentum space. However, this fails because a system of more than one fermion is bound to have a non-zero expectation value for the $V_1$ term. Hence we work in real-space, so that for finite number density the density matrix effectively describes a product state of localized orbitals separated by empty sites. To accelerate convergence, we use the optimal damping algorithm~\cite{Cancs2000}, adapted for three-body interactions. Since the self-consistency loop is easily stuck in local minima for large $V_1$, we initialize multiple seeds, and perform an `annealing' procedure where $V_1$ is slowly varied over an initial range of iterations, ending at $V_1=10^8$.
	
	\section{Fermi liquids and charge order at low fillings $\nu<1/4$}\label{sec:lowfilling}
	
	In the regime of dilute fermion filling, it is expected that local interactions play a subordinate role and the hard-core constraint is largely inactive. As a result, the qualitative behavior of the system is dominated by the hopping processes, which naturally result in the formation of a Fermi surface similar to the one in graphene at high hole doping. The hard-core nature of the fermions and their density-density interactions then enter as corrections to this pure Fermi-gas behavior. By measuring the lattice Green's function $G^{\alpha,\beta}_{i,j}\equiv\langle c\dag_{\beta,j} c\pdag_{\alpha,i} \rangle$, we obtain the momentum-space occupation number $n(\mathbf{k})$ of the ground state wave function as
	\begin{equation}
		n(\mathbf{k}) = \frac{1}{N} \sum_{\alpha=A,B} \sum_{i,j} e^{i \mathbf{k}\cdot(\mathbf{r}_i-\mathbf{r}_j)} G^{\alpha,\alpha}_{i,j}.
	\end{equation}
	For non-interacting fermions on the honeycomb lattice, i.e. without density-density interactions or the hard-core constraint, $n(\mathbf{k})$ should exhibit a clear step (whose magnitude is the quasiparticle residue, $Z$) as a function of the graphene dispersion $\epsilon_{\mathrm{G}}(\mathbf{k})$. \cfig{low_filling}(a) clearly displays a step-like behavior for the occupation at lower fillings $\nu \lesssim 0.2$, consistent with the formation of a Fermi liquid. Increasing the fermion density leads to a gradual softening of the quasiparticle residue,
	so that the Fermi liquid appears to give way to a non-Fermi-liquid phase at $\nu \gtrsim 0.25$.
	
	Interestingly, we find that the Fermi liquid regime is interrupted by a charge ordered phase at filling $\nu=1/6$, which triples the unit cell and spontaneously polarizes into either the $A$ or $B$ sublattice. The expected six-fold ground state degeneracy in the many-body spectrum as well as the clear energetic preference of simulation clusters supporting this form of order strongly points at spontaneous symmetry breaking as the root of this incompressible phase. Additionally, the static structure factor $S_{\mathrm{C}}(\mathbf{q})$ obtained from Fourier transformation of the measured density-density correlation function
	\begin{equation}
		\label{eq:correlation_function}
		C_{i}^{\alpha,\beta}(a) \equiv \langle n_i^{\alpha} n_{i+a}^{\beta} \rangle, \quad \mathrm{where} \quad \alpha=A,B,
	\end{equation}
	shown in \cfig{low_filling}(b), is expected to exhibit pronounced peaks. Indeed, as can be seen in \cfig{low_filling}(c), the peak at $\mathbf{q}=\mathbf{K}$ extrapolates to finite values in the TDL while other signals vanish, indicative of long-range order.
	
	These ED results are corroborated by our HF simulations, where the same type of symmetry breaking order prevails at $\nu=1/6$ and ground state energies per site are relatively close to the ones obtained from ED (cf. \cfig{phase_diagram_schematic} for $\nu\leq 1/6$). HF also finds a charge density wave at $\nu=1/8$ consisting of a fermion delocalized around a honeycomb for every enlarged $2\times 2$ unit cell. This suggests a potentially more general instability towards charge order above some critical filling in the dilute fermion regime.
	
	\section{Zero-energy state window $1/4 \leq \nu \lesssim 0.292$}\label{sec:zeroenergy}
	
\begin{figure}[htp]
		\centering
		\includegraphics[width=1\columnwidth]{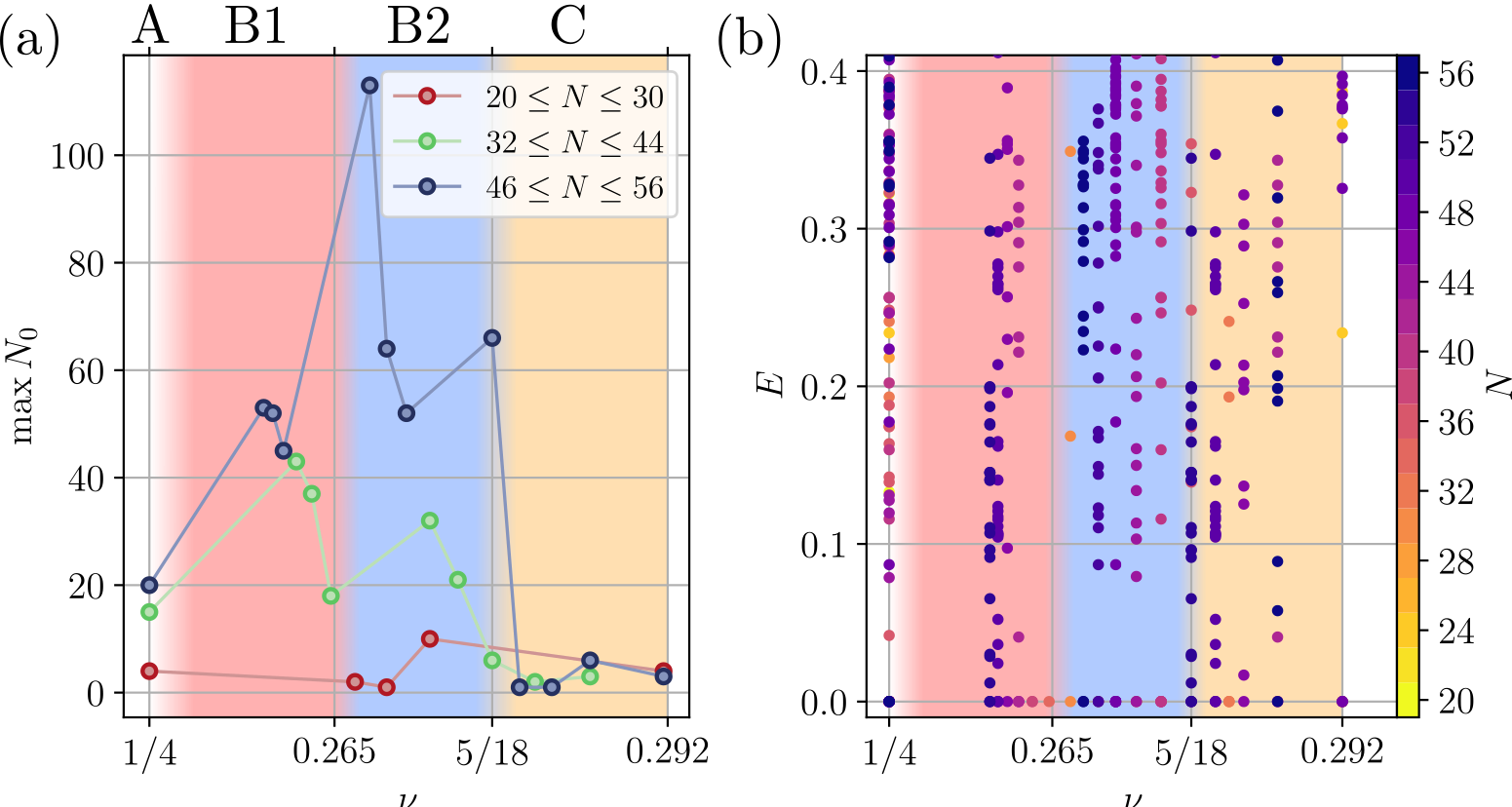}
		\caption{\textbf{Degeneracy and low-energy spectrum in the zero-energy window.} (a) Maximum number of states at zero energy as a function of filling $\nu$ and grouped system sizes $N$. The marked/shaded fillings correspond to the ones discussed in Sec.~\ref{sec:zeroenergy}, where A: $\nu=1/4$; B: $1/4<\nu\leq5/18$, subdivided into (red) B1: $1/4<\nu\leq 0.265$ and (blue) B2: $0.265<\nu\leq 5/18$; as well as (orange) C: $5/18<\nu\lesssim 0.292$. 
			While $N_0$ exhibits clear extensive behavior in A and B, it does not in C.
			(b) By employing a finite-size analysis like \cfig{extensive_entropy_filling}(b), the low-energy spectrum may be used to deduce the gapped nature of A and B2, while B1 appears to be gapless.}
		\label{fig:zero_degeneracy_spectra}
	\end{figure}

	Starting from $\nu=1/4$, i.e.~one fermion per two unit cells, the ground state of the SUSY model on the honeycomb lattice from \ceqn{H_susy_honeycomb} has \textit{exactly} zero energy for a finite range of fillings. In fact, as illustrated in \cfig{zero_degeneracy_spectra}, we find robust zero-energy states for  (almost\footnote{For the two rational fillings accessible to us in $\nu \in (0.286,0.292)$, we suspect zero-energy states emerge for appropriate geometries but were unable to identify them in our study.}) all rational fillings $1/4 \leq \nu \lesssim 0.292$ within our finite-size simulations. This filling window extends beyond the homological predictions of Ref.~\cite{jonsson2010homology}.
	
	In contrast to the finite-energy Fermi-liquid and charge ordered phases discussed in Sec.~\ref{sec:lowfilling}, due to the supersymmetric form of the Hamiltonian in \ceqn{H_susy_general} all these zero-energy states are necessarily singlets of the supercharge operators, and are hence annihilated by $Q$ as well as $Q\dag$. Interestingly, depending on the filling fraction the system exhibits varying degrees of extensivity of the ground state entropy and can be either gapped or gapless.
	
	We now discuss various features observed within this zero-energy filling window.
	\subsection{Resonating charge stripe order at $\nu=1/4$}\label{subsec:linear}
	
\begin{figure}
		\centering
		\includegraphics[width=\linewidth]{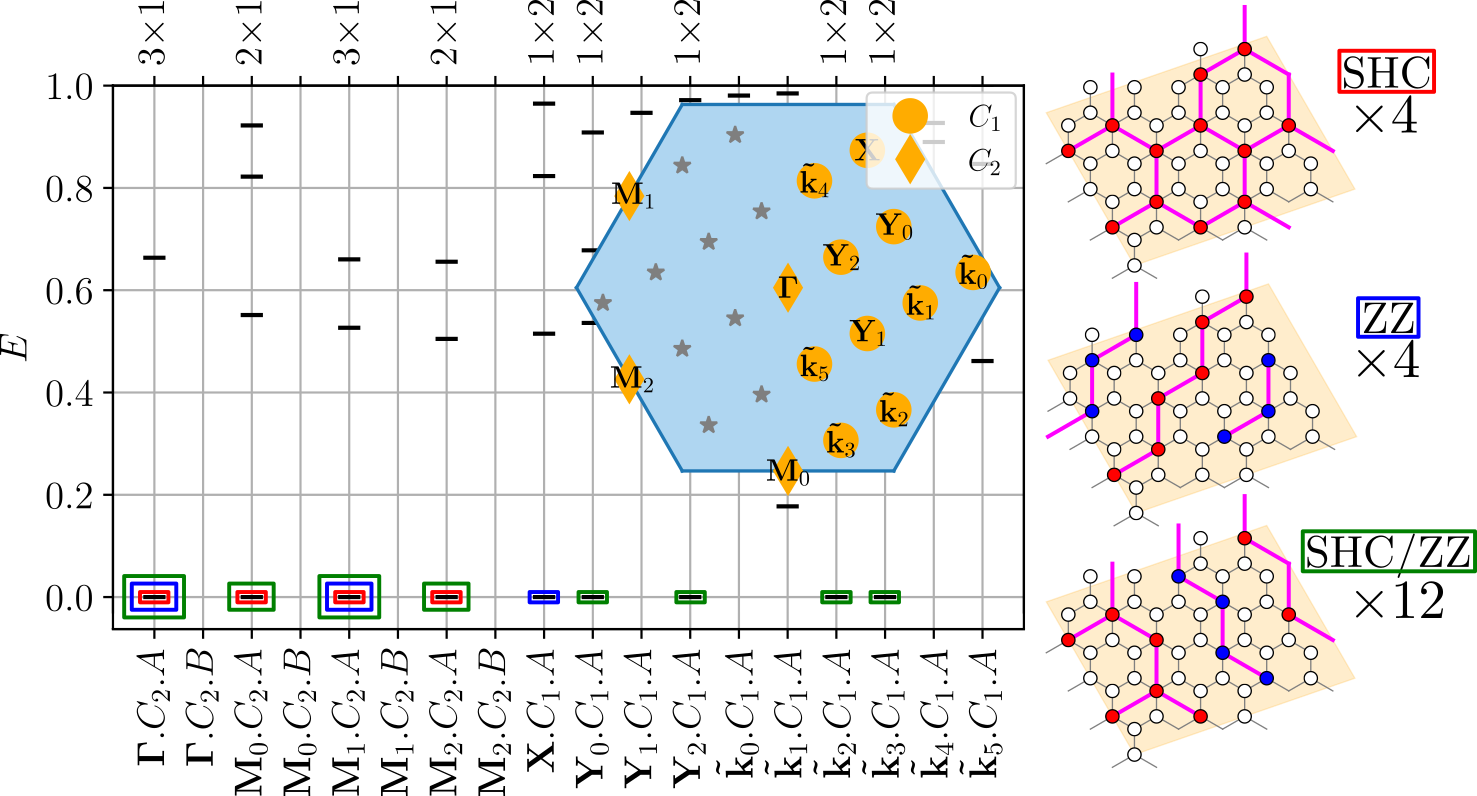}
		\caption{\textbf{Exact diagonalization spectrum at $\nu=1/4$.} Zero-energy states are marked in red. The degeneracy $N_0 = 20$ is a result of spontaneous symmetry breaking in the form of the classical patterns on the right. Besides the more symmetric SHC and ZZ order, commensurate combinations of the two, related by line-sublattice-flips, are also found in the ground state manifold. The lower horizontal axis lists all spatial irreducible representations of the chosen simulation cluster (format is \{trans.-rep.\}.\{little-group\}.\{point-rep.\}) while the upper axis indicates the zero-state degeneracy times the symmetry sector multiplicity. Colored rectangles indicate the specific form of charge order corresponding to each zero-energy level, displayed on the right. 
			The displayed cluster is $\mathbf{u}_1=(2,4),\mathbf{u}_2=(4,-4)$ with $N=48$.}
		\label{fig:quarter_filling_spectrum}
	\end{figure}

\begin{figure}
		\centering
		\includegraphics[width=\linewidth]{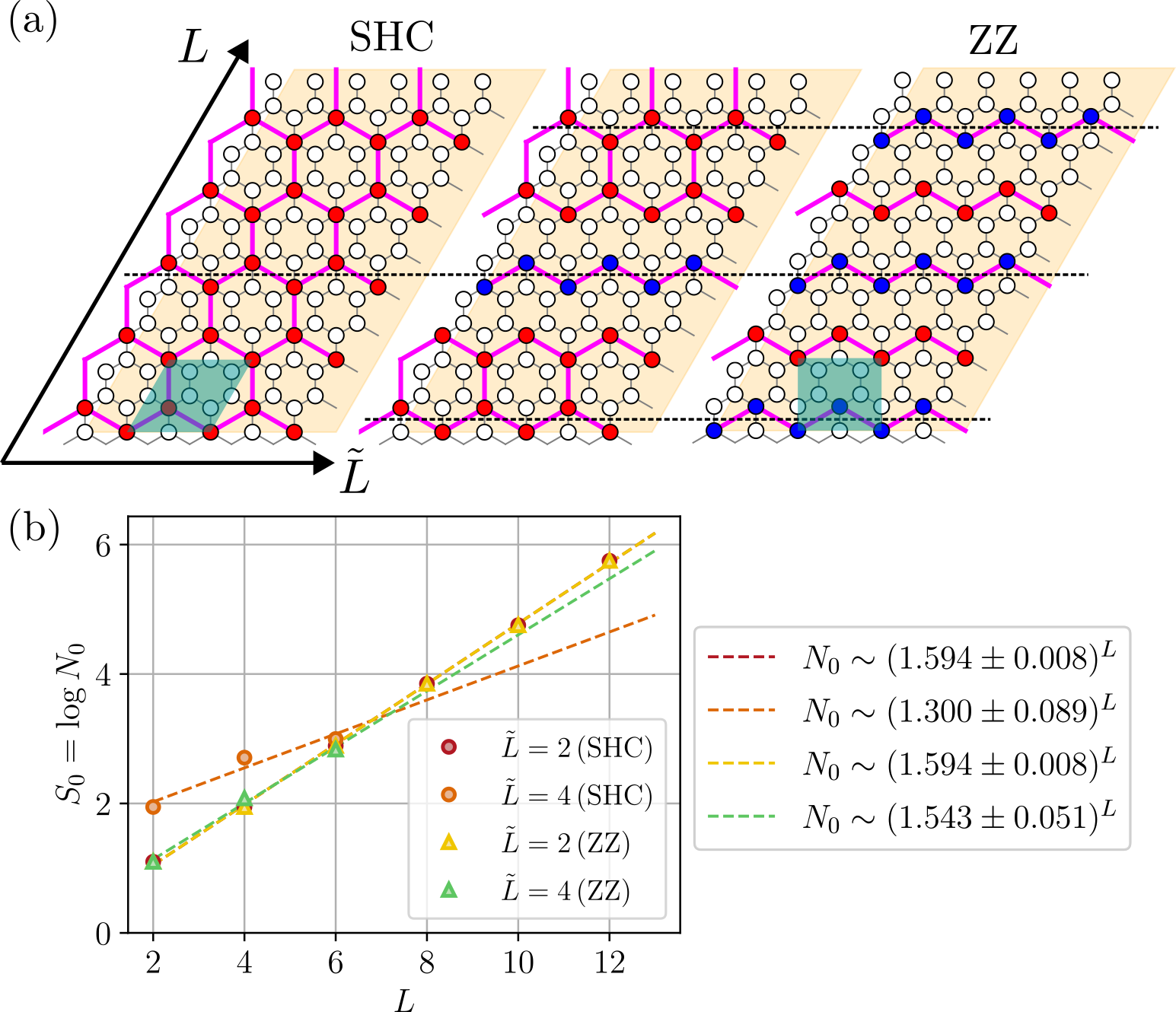}
		\caption{\textbf{Resonating charge stripes at $\nu=1/4$.} (a) Illustration of the generic mechanism of converting between superhoneycomb (SHC) and zig-zag (ZZ) charge order via line-sublattice-flips. These flips along a linear direction do not cost potential energy, such that the wave function can be thought of as \emph{resonating} between different microscopic patterns. (b) The number of states at zero energy increases exponentially in the linear dimension of the torus $L$. The blue shaded ares in (a) indicate the unit cells of SHC and ZZ charge order used to construct the clusters in (b). $\tilde{L}$ denotes the circumference of the torus along the other direction. }
		\label{fig:quarter_filling}
	\end{figure}

	A natural starting point to engineer zero-states at a particular filling is to consider the constrained hopping and interaction parts of \ceqn{H_susy_honeycomb} separately. 
	We first focus on the interaction term, which has both attractive $V_\Delta=-1$ and repulsive $V_2=1$ components. However the net interaction energy is positive semi-definite, because the presence of three particles that activate a $V_\Delta$ contribution necessarily incurs a $3V_2$ penalty. Combined with the hard-core constraint, this leads to the conclusion that the closest-packed configuration with zero interaction energy consists of fermions located at third-nearest neighbor positions, i.e. on the opposite site of a honeycomb plaquette. Requiring next that hopping processes are either blocked by the hard-core constraint or penalized by $V_2$, at quarter-filling one arrives at the super-honeycomb (SHC) and zig-zag (ZZ) patterns shown in \cfig{quarter_filling_spectrum} and \cfig{quarter_filling}. $\nu=1/4$ is the maximum filling at which a configuration with zero potential energy can be realized and therefore has a special role among the zero-energy states. The relevance of these classical motifs can be verified by analyzing their symmetry properties and checking against the degeneracies observed in the many-body spectrum. We indeed find the expected four zero-energy levels in their respective symmetry sectors for each of these long-range patterns to be present in our data. However, as can be seen from \cfig{quarter_filling_spectrum}, the manifold of ground states is not saturated by these states alone.
	
	In fact, ignoring boundary conditions, the SHC and ZZ periodic patterns are related by line-sublattice-flips illustrated in \cfig{quarter_filling}(a). Starting from the SHC pattern, one arrives at ZZ by flipping the sublattice along a line in every other `row' of occupied sites while leaving the charge correlation unchanged along the horizontal direction. Due to the fine balance of terms in the supersymmetric Hamiltonian, these operations and the associated introduced domain walls and `transitional' configurations do not cost energy even in the presence of hopping.
	An analogous symmetry analysis of these mixed patterns yields the expected degeneracies per symmetry sector observed in the ED spectrum (\cfig{quarter_filling_spectrum}) for the majority of geometrically commensurate clusters, corroborating the dominant role of this unusual form of spontaneous symmetry breaking for this filling fraction.
	
	For large enough systems, entropic arguments render the ground state anisotropic since a generic number of line-sublattice-flips generates a \emph{stripe-like} configuration with a period of $2 \mathbf{a}$ only along one direction. As an arbitrary superposition of these states can \emph{resonate} between states of different numbers of line-sublattice-flips, we dub this phase the resonating, charge stripe order (RCS). In contrast to e.g. the charge ordered phase at $\nu=1/6$, the GSD of the RCS is expected to increase exponentially with the linear size of the system. The analysis of the excitation gap $\Delta_{\mathrm{ex}}$, i.e. the first non-zero eigenvalue across all symmetry sectors, as a function of system size $N$ suggests the RCS to be incompressible, consistent with the opening of a gap due to the spontaneous breaking of discrete translational and rotational symmetries.
	Due to the limited system sizes that can be accessed with ED, the less symmetric combinations of SHC and ZZ patterns can only be observed in a handful of simulation clusters with aspect ratios near unity. 
	To study the zero-energy degeneracy $N_0$ and consequently the associated entropy $S_0 = \log N_0$ as a function of linear dimension of the torus $L$, we restrict our discussion to thin but long tori that are commensurate with SHC and ZZ order. We perform a linear fit $S_0 = L \log g_0$ to extract the scaling factor $\log g_0$, which controls the exponential GSD in the thin-torus limit. While caution is required when interpreting the numerical values in \cfig{quarter_filling}(b) due to possible finite-size effects, for both choices of unit cells (of the SHC/ZZ orders) and both choices torus widths ($\tilde{L} =2, 4$) we find that $g_0$ is significantly larger than one at $g_0 \simeq 1.30 - 1.59$ within the accuracy of the fit, implying $\log g_0 \simeq 0.26 - 0.46$. Furthermore, the values are in the vicinity of the classical estimate of $N_0 \sim 2^{L/2}$ which corresponds to $ \log g_0  = \log \sqrt{2}\simeq 0.35$. We obtain this classical estimate by  arbitrarily choosing whether or not to sublattice-flip each of the $L/2$ horizontal lines of charges from either the fully-ZZ or fully-SHC parent state.
	This scaling is consistent with our understanding of a sub-extensive GSE of the RCS.
	
	Although a discussion on the level of classical configurations works well for explaining the majority of observed degeneracies, the Hamiltonian naturally generates quantum fluctuations as the constrained hopping and density-density interactions terms do not commute. While the highest-amplitude basis states have minimal potential energy and reflect the patterns of charge order discussed previously, the hopping processes generate a significant ground state weight onto other configurations, with finite $V_2$.
	This is evident in the clear discrepancy of ground state energies in \cfig{phase_diagram_schematic}, where the mean-field HF approximation, despite converging on one of the correct types of charge order, results in an energy clearly above zero. 
	Therefore, although a classical picture accurately captures the symmetry properties of the RCS phase, it is probable that a significant amount of entanglement and quantum fluctuations is generated for $t\neq 0$.
	
	\subsection{Extensive entropy regime $1/4<\nu \leq 5/18$}\label{subsec:extensive}
	
\begin{figure}
		\centering
		\includegraphics[width=.7\linewidth]{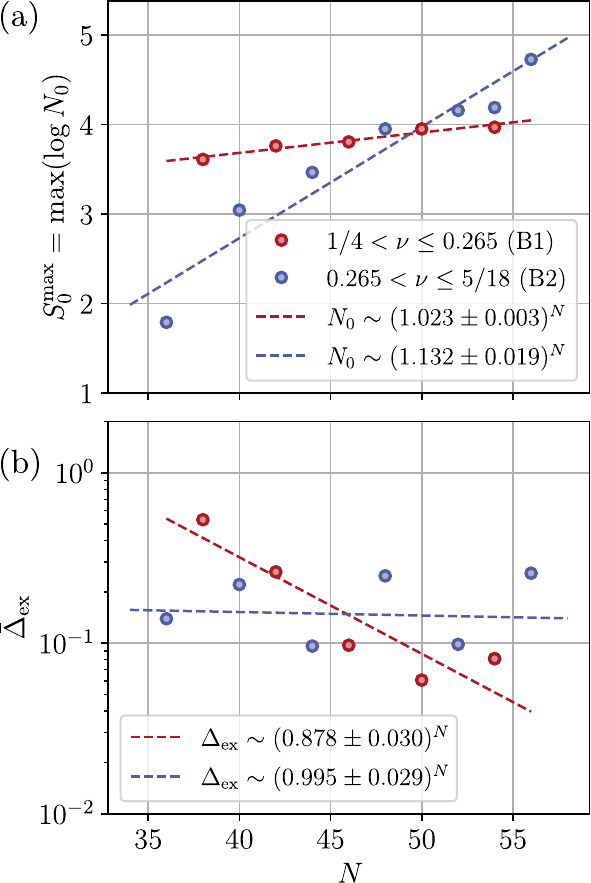}
		\caption{\textbf{Extensive entropy filling window.} (a) The scaling rate $g_0$ of the zero-state-degeneracy $N_0 \sim g_0^N$ differs for the two filling windows B1: $1/4<\nu \leq 0.265$ and B2: $0.265<\nu \leq 5/18$. Since $g_0>1$, the number of ground states grows exponentially with the system size $N$. (b) The gap to the first excited state $\bar{\Delta}_{\mathrm{ex}}$ averaged over all clusters appears to vanish in the TDL for regime B1, consistent with a gapless phase. For B2, the exponent is very close to one, indicating the gap might remain open even for infinite systems.}
		\label{fig:extensive_entropy_filling}
	\end{figure}

	Upon doping additional charge carriers on top of $\nu=1/4$, the condition of minimal potential energy can no longer be exactly satisfied and fluctuations become increasingly important. The model nevertheless realizes zero-energy states whose degeneracy appears to depend non-trivially on both the filling as well as the geometry of the simulation lattice. To study these states, we combine data from multiple nearby fillings that share similar signatures into coarse grained windows. 
	This allows us to subdivide this filling interval into two main, distinct regimes labeled B1: $1/4<\nu \leq 0.265$ and B2: $0.265<\nu \leq 5/18$, as shown in \cfig{zero_degeneracy_spectra}. Studying the maximum zero-state degeneracy $N_0$ as a function of the system size in \cfig{extensive_entropy_filling}(a), we find both intervals to be consistent with an exponential dependence on $N$, implying that the GSE $S_0 \sim N$ is extensive in the area and the system is \emph{superfrustrated}. This property was previously linked to quantum criticality and the implied competition of multiple distinct states of matter \cite{huijse2008susy,Huijse2012triangular,chepiga2021ladder}.
	
	\cfig{extensive_entropy_filling}(a) shows that while $N_0$ is exponential in $N$ for both B1 and B2, their rates differ significantly. While $N_0 \sim 1.02^N$ in B1, $N_0 \sim 1.13^N$ in B2,  with a potential maximum in the vicinity of $\nu \simeq 0.268$ ($N_0=113$ for $N=56$). Hence, for large enough systems, B2 gives the dominant contribution to the total GSD across all filling sectors:
	although on the lower side, the scaling rate of $N_0$ in B2 is consistent with calculations of the Witten index $\abs{W} \sim (1.2 \pm 0.1)^N$ for the same model in Ref.~\cite{vaneerten2005witten}, which provides a lower bound to the total number of states at zero energy across all charge sectors.
	The largest filling fraction exhibiting degeneracies within a symmetry sector\footnote{Such degeneracies are a prerequisite for an exponentially large $N_0$: the degeneracies arising due to {\it distinct} symmetry sectors is at most algebraic as it is bounded by the growth in the number of irreducible representations.} is $\nu=5/18$, coinciding with the upper limit of the cross-cycle construction of Ref.~\cite{jonsson2010homology}. 
	The maximum ground state entropy per site in our data according to the standard definition $S_0^{\mathrm{site}} \equiv \log(N_0)/N$ is located in B1 near $\nu \simeq 0.263$ (for $24<N\leq 56$). This appears to disagree with \cfig{extensive_entropy_filling}(a), where the larger scaling rate $g_0$ is found for B2. The discrepancy is resolved by realizing that if $N_0 = \eta g_0^N$, then $S_0 = \log(\eta) + \log(g_0) N$ and hence $S_0^{\mathrm{site}} = \log(\eta)/N + \log(g_0)$, where according to \cfig{extensive_entropy_filling}(a) the intercept and hence $\eta$ appears to be significantly larger for B1. For large enough systems $\log(\eta)/N$ tends to zero and $S_0^{\mathrm{site}} \simeq \log(g_0)$ such that both analyses coincide, whereas $\log(\eta)/N$ can significantly contribute to $S_0^{\mathrm{site}}$ for the system sizes ($N \leq 56$) in our study. We hence conjecture that for large enough system sizes the maximum of $S_0^{\mathrm{site}}$ is located in B2. The rate obtained from the slope in \cfig{extensive_entropy_filling}(a) corresponds to an entropy per site of $S_0^{\mathrm{site}} \simeq 0.124$, which is lower than the best estimate for the triangular lattice at $S_0^{\mathrm{site}} \simeq 0.152$ \cite{galanakis2012triangular} and explains why comparatively large systems are required to uncover the extensive nature of the GSE in numerics.
	
	Besides their difference in the scaling of the GSE, the filling intervals B1 and B2 seems to show contrasting behavior in their compressibility. The spectral gap $\Delta_{\mathrm{ex}}$ displayed in \cfig{extensive_entropy_filling}(b) decays with system size for fillings in B1. In particular, the gap appears to close exponentially, $\Delta_{\mathrm{ex}} \lesssim e^{- 0.128 N}$. This is possibly indicative of a spontaneously broken discrete symmetry, although we are not able to identify the relevant symmetry given that we do not understand the `classical' trial state at these fillings. In contrast, the data for $\Delta_{\mathrm{ex}}$ in B2 oscillates with system size. Importantly, the mean $\Delta_{\mathrm{ex}}$ of the oscillation does not seem to decay but, within the accuracy of our fit, remains constant with $N$, indicating the persistence of a finite gap in the thermodynamic limit.
	Introducing a small third-neighbor interaction $V_3$, we observe that remnants of the $\nu=1/4$ motifs are present in the dominant Fock-space configurations of some clusters, pointing to the possibility that some of the extensive entropy states might originate from introducing additional charge carriers into the already sub-exponentially degenerate RCS phase.
	
	\subsection{High filling zero-energy states $5/18 < \nu \lesssim 0.292$}\label{subsec:coincident}
	
	Starting from $\nu>5/18$, we find the ground state to be unique in its symmetry sector. Although degeneracies across different irreducible representations are still present, the total number of zero-energy states due to these can only grow algebraically and hence the GSE cannot be extensive. 
	The reduction of degeneracies towards the right end of the zero-energy window is a robust feature observed across multiple system sizes, which is intuitively understood by the increased importance of the hard-core constraint and the net repulsive density-density interaction for higher fillings and is consistent with analogous observations on the triangular lattice \cite{galanakis2012triangular}.
	Up to $\nu \simeq 0.286$, all fillings realizable in our finite size systems have a finite number of states at zero energy. Furthermore, the highest filling with robust $N_0>0$ is $\nu \simeq 0.292$, being realized by multiple relatively large clusters up to $N=48$.
	For $\nu \simeq 0.288$ and $\nu \simeq 0.289$, no zero-energy states are found in our simulations. We believe this is a consequence of the more stringent geometric conditions for vanishing energy at higher fillings, which are not satisfied by the clusters used in this study. 
	We conjecture that for the SUSY model on the honeycomb lattice in the TDL, all fillings up to at least $0.292$ have $N_0>0$, significantly extending the width of the zero-energy window by a factor of $\sim 1.51$ compared to Ref.~\cite{jonsson2010homology}. 
	The ED spectra suggest a tendency for the excitation gaps $\Delta_{\mathrm{ex}}$ to decrease with system size, though a definite conclusion about the compressibility in this regime would be too speculative given the available data. 
	
	\begin{figure}
		\centering
		\includegraphics[width=\linewidth]{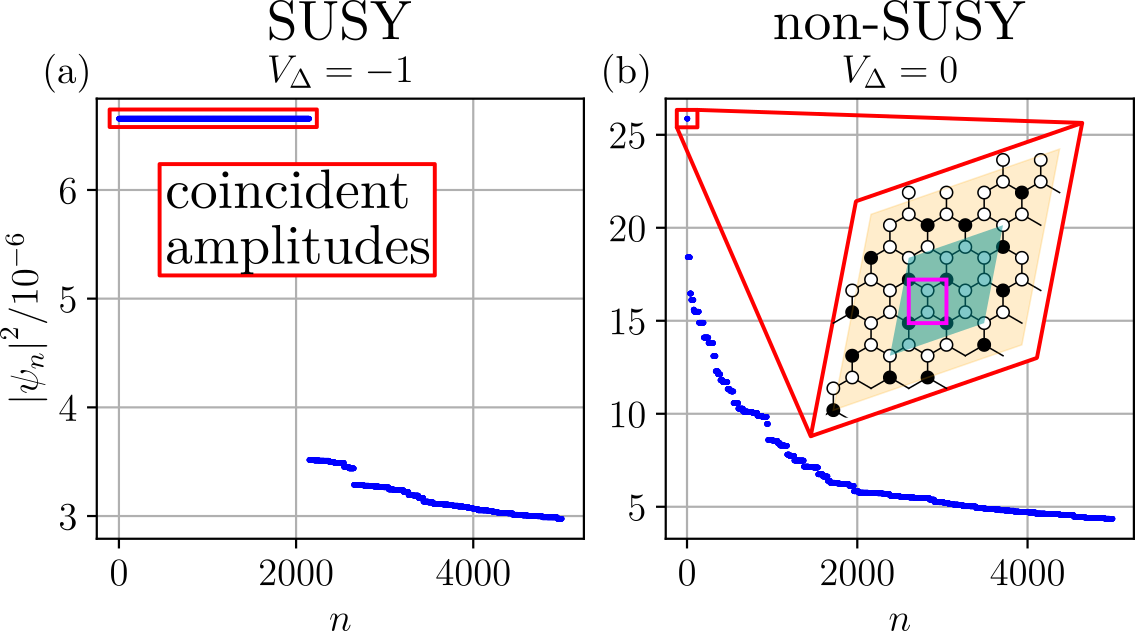}
		\caption{\textbf{Coincident amplitudes at $\nu=2/7$.} (a) Amplitudes of the leading real-space configurations in the ideal SUSY limit at $\nu=2/7$ exhibiting the coincident amplitude feature. All degenerate configurations have the same, minimal classical potential energy but vary in their $V_2$ and $V_{\Delta}$ contribution. (b) The feature is unstable towards perturbations in the couplings of \ceqn{H_susy_honeycomb}, sub-selecting a particular type of configuration from the previously degenerate manifold and potentially resulting in a more classical state. The highlighted `box'-like charge order is also present in the manifold of coincident amplitudes.
			The used cluster is $\mathbf{u}_1=(2,4),\mathbf{u}_2=(6,-2)$ with $N=56$. }
		\label{fig:coincident_amplitude}
	\end{figure}
	
	Furthermore, we observe the presence of a curious `coincident amplitude' feature. It is defined by a substantial (numerically) \emph{exact} degeneracy of the leading amplitude in the computational basis of the many-body wave function, and was also reported in Ref.~\cite{galanakis2012triangular} for high filling zero-energy states in the triangular lattice SUSY model. 
	We find these signatures to be universally present for large enough systems ($N>20$) in this filling regime.
	\cfig{coincident_amplitude}(a) shows a particularly striking instance, where more than two thousand distinct (81 symmetry-inequivalent) real-space configurations constitute the leading set of coincident amplitudes. The degree of this degeneracy varies with both system size and geometry, which prevents a more in-depth understanding of the origin of this feature. 
	Tuning away from the SUSY limit in \cfig{coincident_amplitude}(b) verifies this property to be related to the fine-tuned nature of the couplings $t,V_2,V_{\Delta}$ entering the extended Hubbard-like model of \ceqn{H_susy_honeycomb}, rather than the operator structure itself. In the perturbed case of \cfig{coincident_amplitude}(b) at $\nu=2/7$ with $V_{\Delta}=0$, a box-like charge order pattern emerges over the rest of the degenerate manifold as the highest-amplitude configuration.
	The coincident amplitudes appear to not be exclusive to this high filling regime, as we also observe instances of a significant degeneracy at $\nu=5/18$ as well as to a minor degree on a selected cluster at $\nu \simeq 0.268$ with a finite energy ground state. The latter example appears to indicate that coincident amplitudes are not necessarily tied to the annihilation by the supercharge operators $Q$ and $Q\dag$. However, we do not observe comparable signatures outside the zero-energy window. Attaining a more general understanding of its relevance is hindered by the degeneracies within a symmetry sector present for $1/4 \leq \nu \leq 5/18$, where such a feature is no longer well-defined. 
	We also investigated the possibility of explicitly constructing a ground state by exploiting the manifold of coincident amplitudes, but we found the weights across all basis states\footnote{On smaller systems of $N \sim 20 - 30$, where we measured the weights of the full wave function.} to decay over multiple orders of magnitude, which challenges a `simple' ansatz for the wave function. Still, the coincident amplitude feature could guide a more optimal choice of basis which could facilitate a more analytical understanding.
	
	\section{Sublattice bubbles and domain walls at high filling $0.292 \lesssim \nu\lesssim1/2$}
	
\begin{figure}[htp]
		\centering
		\includegraphics[width=.8\columnwidth]{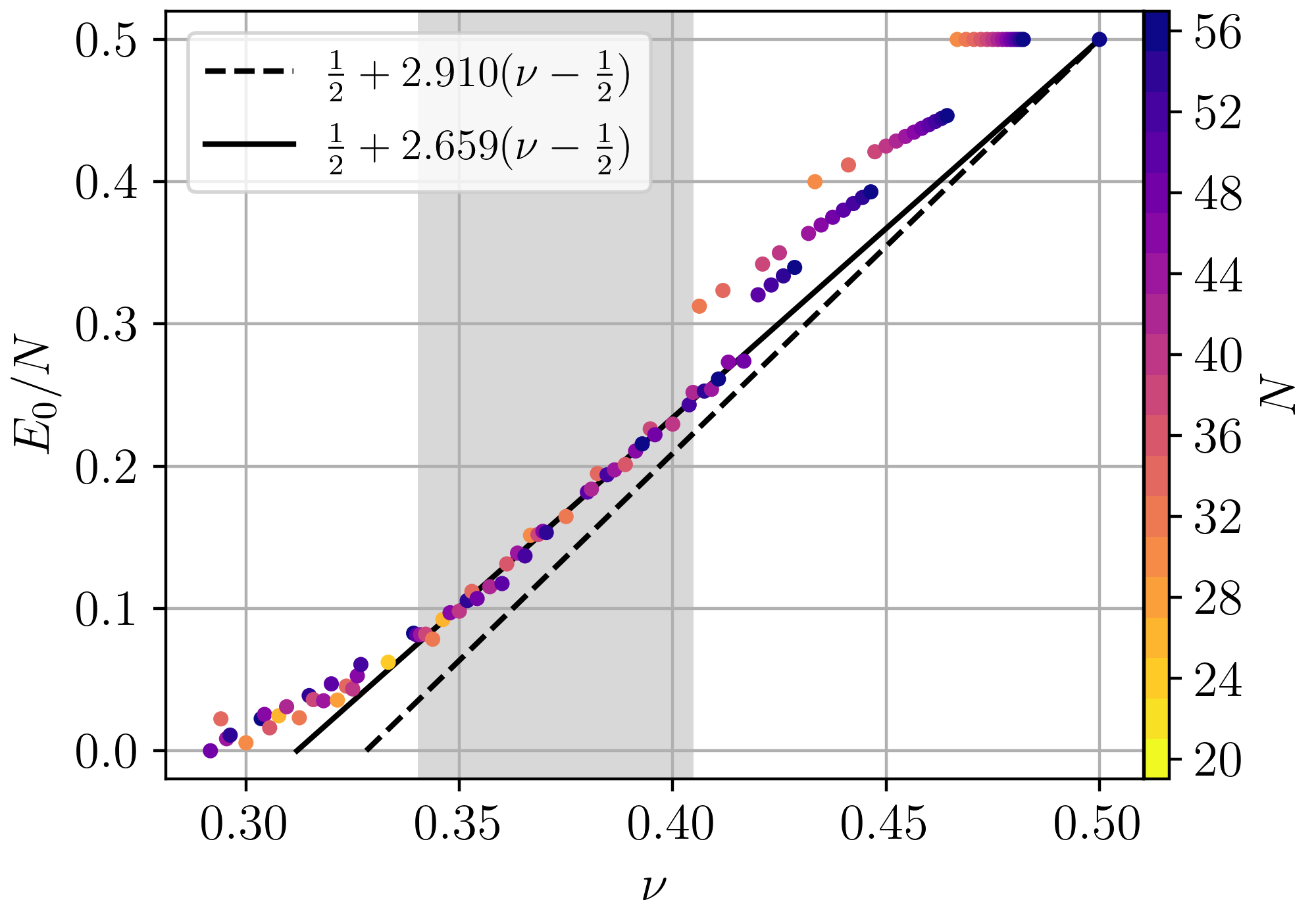}
		\caption{\textbf{Hole-doping energies per site.} Energies per site in the dominantly hole-doped regime near $\nu=1/2$. Dashed (solid) lines correspond to slopes $\alpha$ obtained by mapping to the XXZ chain $\alpha = (1+6/\pi) \simeq 2.91$ (fitting to the ED data in the gray shaded interval $\alpha=2.659(5)$). The four fan-like structures in $E_0/N$ near the right end of the filling axis correspond to sublattice bubbles, whose critical filling tends to $\nu=1/2$ in the TDL.}
		\label{fig:hole_doping_energy}
	\end{figure}

	At one fermion per unit cell, corresponding to $\nu=1/2$, the Hilbert space of the problem is maximally reduced, and consists of just the pair of states that are fully polarized on either of the two sublattice $A$ or $B$, with energy $E_{\frac{1}{2}} = N/2$. Upon doping with a finite number of holes,  we find that finite systems initially realize a `bubble' of opposite sublattice polarization embedded in the inert $\nu=1/2$ crystal. This type of bubble formation is the origin of the fan-like signatures in $E_0/N$ in \cfig{hole_doping_energy}. As a result of the sublinear scaling of charge with system size (as it is proportional to the area of the bubble), the critical hole filling tends to zero as $N \rightarrow \infty$ and the physics is likely to be dominated by sublattice domain walls with (in the nearly free case) linear $E$ vs. $\nu$. The fluctuations (and eventually interactions) of these domain walls increase towards $\nu \rightarrow 0.292$, resulting in the convex bend in \cfig{hole_doping_energy} near $\nu \simeq 0.3$, which appears to smoothly connect to zero energy. In these filling regimes ($0.3 \lesssim \nu \lesssim 0.35$ as well as $0.35 \lesssim \nu \lesssim 0.4$), the analysis of excitation gaps in the many-body spectrum points to a compressible, potentially liquid-like phase.
	
	\label{sec:highfilling}
	
	\begin{figure*}
		\centering
		\includegraphics[width=0.85\linewidth]{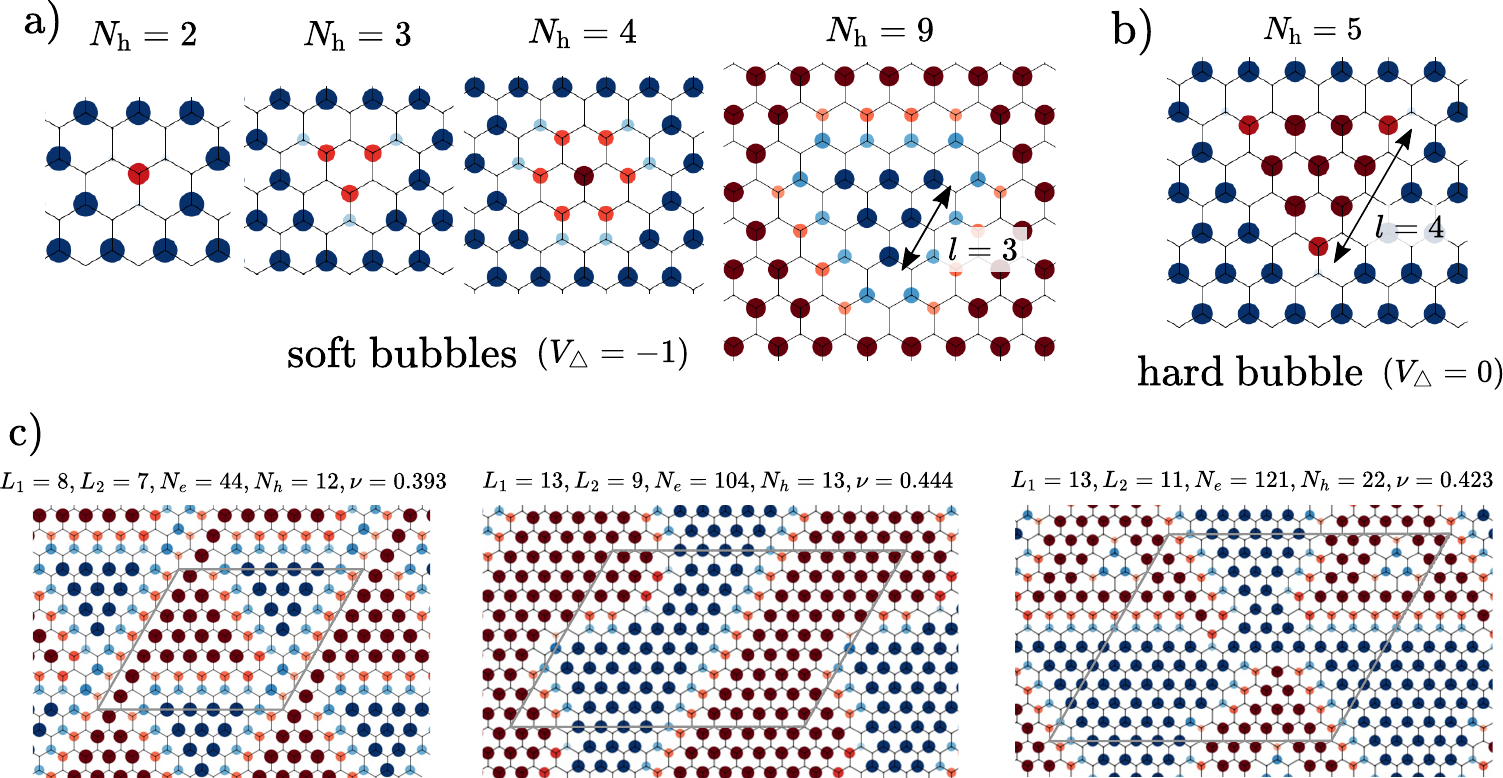}
		\caption{\textbf{Sublattice polarization bubbles and domain walls at $\nu\lesssim1/2$ in Hartree-Fock.} a) `Soft' bubbles formed upon doping the maximally filled state at $\nu=1/2$ with $N_\text{h}$ holes. b) Representative `hard' bubble away from the SUSY limit ($V_\triangle=0$). c) Example domain wall configurations for larger hole dopings in the SUSY limit.}
		\label{fig:bubbles}
	\end{figure*}
	
	\subsection{Mean-field analysis of bubbles}\label{subsubsec:MFbubbles}
	
	For small hole-doping from $\nu=1/2$, the system is strongly blockaded, and the highly constrained nature of the problem suggests that a mean-field analysis may perform well. Indeed, numerical HF energies are comparable to ED near half-filling (Fig.~\ref{fig:phase_diagram_schematic}). In this section, we first consider a small number of holes $N_\text{h}$ doped on top of the sublattice-polarized state, and then consider a family of mean-field bubble ans\"atze of arbitrary size (Fig.~\ref{fig:bubbles}a). Throughout, we assume that the system size is large enough so that boundary effects do not occur.
	
	As mentioned above Hilbert space at $\nu=1/2$ comprises two states, where one of the sublattices is fully filled; evidently these are real-space Slater determinants, i.e. mean-field theory is exact at $\nu=1/2$. 
	The exact ground states for $N_\text{h}=1,2$ can also be captured by single Slater determinants, as we now describe. 
	
	Starting from the $A$-polarized state, the 1-hole subspace is $N/2$-dimensional, consisting of configurations where the fermion on one site has been removed. The resulting basis states are \emph{frozen} as the kinetic term is still unable to hop any fermions, leading to degenerate states with $E_{1\text{h}}=E_{\frac{1}{2}}$.
	
	For $N_\text{h}=2$, the fermions that we remove are either nearest neighbors or further apart. The change in interaction energy in both cases is $-6$. However, only the former case is relevant for the ground state since one of the remaining fermions is able to delocalize around a tripod (Fig.~\ref{fig:bubbles}a), and hence benefit from the kinetic term, which is impossible unless the doped holes are adjacent. Note that the rest of the system is still frozen into a sublattice-polarized configuration. The effective single-particle hopping model for the mobile fermion can be solved to obtain $E_\text{2h}=E_{\frac{1}{2}}-3$, i.e.~$-\frac{3}{2}$ per hole.
	
	For $N_\text{h}=3$, we can remove three fermions around a hexagon, which allows three other fermions to hop around a triangular region of nine sites (Fig.~\ref{fig:bubbles}a). This is now a genuinely interacting few-body problem, so that the mean-field estimate  $E^\text{MF}_{3\text{h}}\simeq E_{\frac{1}{2}}-5.50$ is only an upper bound for the ground state energy, which is ~$-1.83$ per hole. However, solving the three-body problem exactly recovers the linear fan of $N_\text{h}=3$ ED energies. An analogous procedure also gives the exact ED energies for $N_\text{h}=4$. These observations reinforce the hypothesis that the linear fans near $\nu=1/2$ correspond to single `bubbles' of the half-filled state for cluster geometries where the periodic boundaries do not yet play a role. These states can be created by emptying out the fermions in a triangular region (upward pointing for the $A$ sublattice), and then reintroducing a smaller number of fermions into the same region, while the rest of the system remains frozen and dynamically decoupled. This construction underlies the Hilbert space fragmentation that occurs more generally for hard-core extended Hubbard models with n.n. hoppings, discussed elsewhere~\cite{Kwan2023hardcore}.
	
	The HF configurations for larger $N_\text{h}$ suggest that the bubble consists of a closed triangular domain wall loop that separates bulk regions of opposite sublattice polarization. Domain walls can run along any of three directions orthogonal to the honeycomb bonds, but they are `soft' or `hard' depending on the direction and the sublattice polarization of the adjacent domains (Fig.~\ref{fig:bubbles}a,b). If we assign directions such that the $A$ sublattice (blue) is on the left when running along the domain wall, then the soft domain wall corresponds to angles $0,\frac{2\pi}{3},\frac{4\pi}{3}$ relative to the $x$-axis, while the hard domain wall corresponds to $\frac{\pi}{3},\pi,\frac{5\pi}{3}$. The electrons at soft domain walls are able to delocalize, but the ones at hard domain walls are localized on a single site because of the hard-core constraint. Below, we work with general Hamiltonian parameters $t,V_2,V_\triangle$ to understand the competition between the two domain wall/bubble types.
	
	At mean-field level, the soft bubble consists of inner and outer polarized bulks, and a delocalized triangular ring of fermions. Such a bubble of linear size $l$ contains $2l+3$ holes spread on its boundary. For simplicity, we consider the ansatz where all the boundary fermions are fully delocalized across the two sites they have access to. While there are corrections arising from the corner regions, they have a vanishing contribution for large $l$. The (negative) energy of such a bubble is
	\begin{equation}
		\begin{aligned}
			\Delta E^{\text{MF}}_{\text{soft}}(l)&=|t|\left[-3l-3\right]+V_2\left[-\frac{27}{2}l-\frac{81}{4}\right]
			\\&+V_\Delta\left[-\frac{11}{2}l-15\right]+3\left[2l+3\right]
		\end{aligned}
	\end{equation}
	Hence for large $l$, the energy per hole is $\Delta\epsilon^{\text{MF}}_\text{soft}(l\rightarrow\infty)=-\frac{3}{2}|t|-\frac{27}{4}V_2-\frac{11}{4}V_\Delta+3$. 
	
	A similar computation for the hard bubble of linear size $l$ with $l+1$ holes leads to
	\begin{equation}
		\begin{aligned}
			\Delta E^{\text{MF}}_{\text{hard}}(l)&=V_2\left[-9L-6\right]+V_\Delta\left[-5l-2\right]+3\left[l+1\right],
		\end{aligned}
	\end{equation}
	where we have assumed that the corner fermions are not delocalized. This gives $\Delta\epsilon^{\text{MF}}_\text{hard}(l\rightarrow\infty)=-9V_2-5V_\Delta+3$. Note that the absence of any dependence on the hopping $t$ reflects the frozen nature of the hard domain wall. We have that  $\Delta\epsilon^{\text{MF}}_\text{soft}(l\rightarrow\infty)>\Delta\epsilon^{\text{MF}}_\text{hard}(l\rightarrow\infty)$ in the SUSY limit, consistent with the observation of soft bubbles in the HF numerics.
	
	An important question is whether such bubbles are stable for large $l$. It may happen that it is energetically preferable to roughen the domain walls, or split into multiple smaller bubbles. On general grounds, we know that the relative energy (compared to $E_{\frac{1}{2}}$) and number of holes of a bubble of linear size $l$ is $\Delta E(l)=-\alpha l - \beta$ and $h(l)=\gamma l+\delta$. 
	The relevant quantity is the energy density (per hole) $\Delta E/h=-\frac{\alpha}{\gamma}+\left(\frac{\delta\alpha}{\gamma}-\beta\right)/h$, which needs to be monotonically decreasing for stability against dissociation into smaller bubbles. For the soft bubble in the SUSY limit, the $1/h$ coefficient is positive, reflecting stability against dissociation. This suggests that the bubble corners are less energetically favourable than the straight segments. Hence at mean-field level, added holes to the sublattice polarized state will enter as one large triangular bubble. Since the charge of the bubble is sublinear in its enclosed area, in the TDL, this single-bubble `phase' therefore occupies a vanishing fraction of the filling axis.  
	
	\subsection{Domain wall arrays and phase separation}
	
	An important quantity for understanding the phase structure for the high filling regime in the TDL is the (negative) domain wall tension per hole. In mean-field theory, this is simply given by the $l\rightarrow\infty$ limit of the soft bubble, so that the energy per hole is $\epsilon_{\text{DW}}^{\text{HF}}=-2.5$. As reviewed in App.~\ref{app:exactXX}, the exact line tension for a single isolated domain wall can be computed by mapping to an XX spin chain~\cite{Mila1994,Henley2001stripes,Zhang2003stripes}, leading to an energy per hole of $\epsilon_{\text{DW}}=-(1+\frac{6|t|}{\pi})\simeq -2.91$. In the spin language, the mean-field result corresponds to the case where every second exchange interaction is switched off.
	
	At first sight, the envelope of the ground state energy in Fig.~\ref{fig:hole_doping_energy} appears concave at $\nu\gtrsim 0.4$, implying the presence of phase separation. However, the energy density assuming non-interacting domain walls (dashed line), extrapolated from $\nu=1/2$, lies clearly below the ED data, and its slope is steeper than a linear fit to numerical results for $0.34\leq \nu \lesssim 0.4$. Note that this contrasts with the triangular lattice where the situation is less clear-cut~\cite{galanakis2012triangular}. This suggests the potential relevance of domain walls in the TDL, which is obscured in ED due to the finite system sizes (for sufficiently small number of holes, the domain walls fold into bubbles with relatively expensive corner regions). How does the domain wall physics at $\nu\lesssim1/2$ connect to the dense liquid phase at $\nu\gtrsim 0.3$? There are three scenarios to consider~\cite{Henley2001stripes,Zhang2003stripes,Zhang2004dilute,galanakis2012triangular}:
	\begin{enumerate}[label=(\roman*)]
		\item There is a single stable thermodynamic phase. The equation of state is convex, and the domain wall physics smoothly crosses over to the dense liquid as the filling is reduced.
		\item There is a finite filling window $\nu>\nu_{\text{DW}}^*$ where the fluctuating domain wall phase is stable. At $\nu_{\text{DW}}^*$, there is a first-order transition to the dense liquid phase at $\nu<\nu_{
			\text{DL}}^*$.
		\item There is direct phase coexistence between the dense liquid phase at $\nu<\nu_{
			\text{DL}}^*$ and the fully sublattice polarized state at $\nu=1/2$.
	\end{enumerate}
	
	A crucial detail of the domain wall phase is whether its equation of state bends upwards when hole-doping from $\nu=1/2$. If it is convex, then cases (i) and (ii) are likely, but a concave curve would realize (a variant of) case (iii) with phase coexistence between the sublattice polarized state and some other unknown phase. This is related to the nature of the domain walls beyond the non-interacting XX limit of a single isolated domain wall --- i.e., it is controlled by the physics of multiple domain walls that generically interact. For instance, concavity could arise if the domain walls attract each other. In this case, the they would clump together, which is hardly distinguishable from a liquid-like droplet. We believe this is unlikely due to the geometric constraints of the domain walls on the honeycomb, where two parallel soft domain walls are impossible according to the analysis in Sec.~\ref{subsubsec:MFbubbles}. Another condition concerns adding holes to a system in the putative domain wall phase: they should enter as new domain wall segments, rather than merging into the interior of existing ones. The latter suggests droplet formation and phase separation. We do not see evidence of such behavior in the highest-amplitude configurations of the ED data for large $\nu$. Hence, we believe that cases (i) and (ii) are more likely, but it is not possible to make a definitive determination on the basis of the current finite-size numerical data.
	
	\section{Discussion and conclusion}\label{sec:conclusions}

	As shown in Sec.~\ref{subsec:extensive}, the SUSY model on the honeycomb lattice hosts an extensive ground state entropy regime in the filling interval $1/4<\nu \lesssim 5/18$. Depending on the density of charge carriers, we find the system to be gapless for $0.25 < \nu \lesssim 0.265$ or gapped for $0.265 \lesssim \nu \leq 5/18$ with the maximum ground state entropy likely to be located in the vicinity of $\nu \simeq 0.268$. These results are qualitatively different from what was observed in an ED study of the related triangular lattice \cite{galanakis2012triangular}, where upon doping beyond the lower limit of the zero-energy window located at $\nu=1/7$, the system intially retained the gapped, crystal-like character with no extensive entropy. Furthermore, the filling corresponding to maximum entropy per site was found to be in a gapless regime, while for the honeycomb model our data suggests that the analogous filling is gapped.  
	We caution that given the finite-size nature of the current numerics, we cannot exclude the presence of additional filling regimes that could emerge from the two coarse grained windows (B1 and B2) discussed here. 
	
	Investigating in more detail the properties of the zero-energy states in the regime of exponential degeneracy is beyond the scope of this work. Our characterization of the phenomena at the extremes of the zero-energy window likely gives valuable insights along this direction. In particular, a deeper understanding of the wavefunctions for the RCS at $\nu=1/4$ and the coincident amplitude states near $\nu=2/7$ will provide clues regarding the origin of the extensive zero-energy degeneracy at intermediate fillings. A related direction is to explore the phase diagram when tuning the system away from the multicritical SUSY limit, and understand the competing orders that emerge.

	\begin{figure}[htp]
		\centering
		\includegraphics[width=0.9\columnwidth]{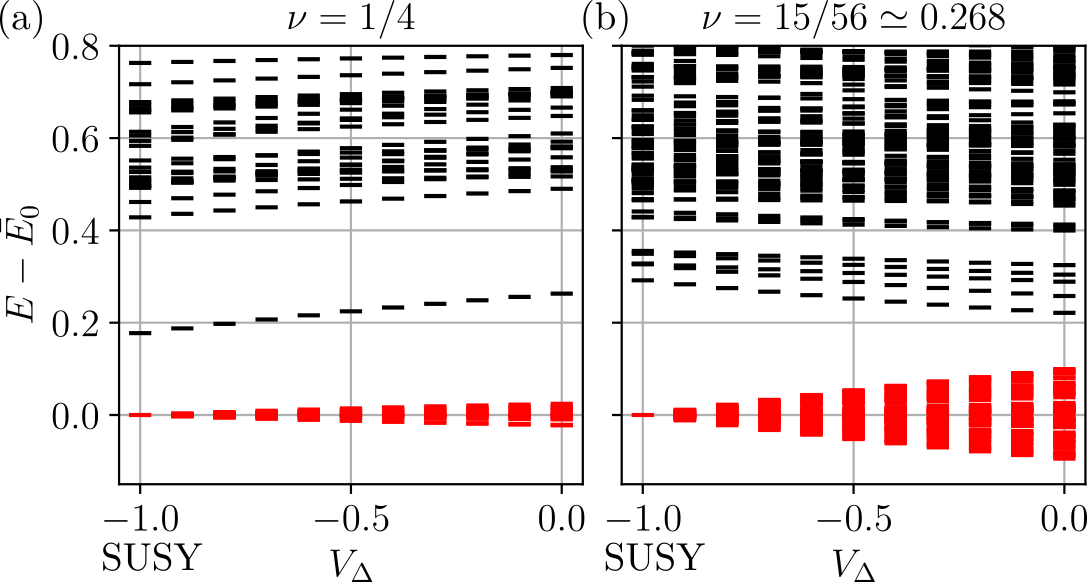}
		\caption{\textbf{Gapped spectra upon tuning down $V_{\Delta}$.} The zero-energy states of the SUSY model (marked in red) split for $\abs{V_{\Delta}}<1$, but remain energetically separated from higher excited states all the way to $V_{\Delta}=0$ for both (a) the RCS phase at $\nu=1/4$ as well as (b) the gapped extensive entropy states at $\nu \simeq 0.268$. This suggests that certain features of the SUSY model could also be observed in experiments implementing the simpler model with two-body terms only. 
			The used cluster in (a) is $\mathbf{u}_1=(2,4),\mathbf{u}_2=(4,-4)$ with $N=48$ and $N_0=20$ while in (b) we use $\mathbf{u}_1=(1,5),\mathbf{u}_2=(5,-3)$ with $N=56$ and $N_0=113$. $\bar{E}_0$ is the median of the $N_0$ lowest energies.}
		\label{fig:tune_Vtri}
	\end{figure}
	
	In Fig.~\ref{fig:tune_Vtri}, we show the many-body spectrum for various gapped fillings in the zero-energy window as a function of the three-body interaction $V_\Delta$. The two fillings correspond to the RCS phase and the exponentially degenerate regime, and represent qualitatively different physics. While the manifold of zero-energy states splits as $V_\Delta$ is detuned, the excitation gap between this manifold and higher-energy states remains open all the way to $V_\Delta=0$ where the Hamiltonian consists just of one-body and two-body terms. This suggests that the physics of the zero-energy states in the SUSY model persists at low energies even in the simpler model where the only interactions are two-body. This is potentially relevant for experimental implementations that can realize extended variants of the Hubbard model, such as Rydberg-dressed atoms in cold optical lattices. Interference of laser beams generates the requisite lattice and hopping processes $t$, while the distance dependence of the interaction potential can be tuned~\cite{Henkel2010rydberg,Pupillo2010rydberg,Johnson2010rydberg} to mimic the extended hard-core constraint and a finite next-nearest neighbor interaction $V_2$~\cite{minar2022rydberg}. There have also been proposals to engineer three-body interactions in such platforms~\cite{Bchler2007three,Gambetta2020engineering,MyersonJain2022fractal}. 
	
	This hierarchy of scales is also potentially relevant for strongly-interacting moir\'e systems. For example, the interaction strength and interaction-renormalized bandwidth of a flavor-polarized band in twisted bilayer graphene are comparable. Furthermore, the Wannier orbitals are centered on a moir\'e honeycomb lattice, and their extended spatial profiles imply strong higher-neighbor interactions. In Refs.~\cite{Zhang2022fractional,Mao2022fractionalization}, this was captured by a model Hamiltonian whose lowest-order form consists of $t\rightarrow 0$ and extended hard-core constraints $V_{1},V_2,V_3\rightarrow\infty$, which was shown to realized correlated insulators at fractional filling and excitations with restricted mobility. We believe that the FSX honeycomb model with $t=V_2=-V_\Delta$ and $V_1\rightarrow\infty$ realizes a complementary limit which may provide insights into the correlation physics of twisted bilayer graphene and related moir\'e heterostructures with a hexagonal moir\'e lattice and appreciable higher-range interactions.
	
	Finally, it may be illuminating to relate the physics at intermediate and high fillings to the Hilbert space fragmentation that was discovered in Ref.~\cite{Kwan2023hardcore} for a more general class of hard-core interacting models. While such fragmentation was shown to be weak for the honeycomb and square lattices and absent for the triangular lattice, other geometries such as the Lieb lattice are strongly fragmented, implying an exponential shattering of Hilbert space connectivity across all fillings in the thermodynamic limit. It would be interesting to investigate how the breakdown of ergodicity in such systems interplays with the enlarged ground state degeneracy expected from the SUSY Hamiltonian.
	
	\begin{acknowledgements}
		We thank Paul Fendley for useful discussions and careful reading of this manuscript.
		We acknowledge support from the European Research Council (ERC) under the European Union Horizon 2020 Research and Innovation Programme (Grant Agreement Nos.~804213-TMCS) and the Austrian Science Fund FWF within the DK-ALM (W1259-N27). The computational results presented have been achieved in part using the Vienna Scientific Cluster (VSC).
	\end{acknowledgements}

	\clearpage
	\newpage
	
	\begin{appendix}
		
		\section{Exact line tension of a domain wall}\label{app:exactXX}
		
\begin{figure}
			\centering
			\includegraphics[width=1\linewidth]{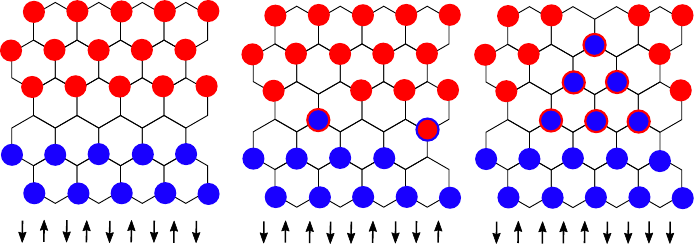}
			\caption{\textbf{Mapping of domain wall to XX chain.} For a horizontal domain wall, the equivalent spin configuration can be determined by considering the relative heights of the boundary fermions belonging to the domain polarized on the A (blue) sublattice.}
			\label{fig:exact_XX}
		\end{figure}

		In this section, we review the calculation of the exact line tension of a single infinite (soft) domain wall on top of the sublattice polarized state. Without loss of generality, we start from a horizontal domain wall that separates semi-infinite bulks of opposite sublattice polarization, with region A below region B. As shown in Fig.~\ref{fig:exact_XX}, the hopping term is able to deform the domain wall.  Following Ref.~\cite{Mila1994,Henley2001stripes,Zhang2003stripes}, we map the fermion configuration to a 1d spin configuration with 2 spins per unit horizontal length of the domain wall. The spins are determined by comparing the relative `heights' of two consecutive columns of A fermions. The hopping term acts non-trivially as an XX coupling term. Note that for the SUSY Hamiltonian where $V_2=-V_\triangle$, the different configurations all have the same potential energy. If $|V_\triangle|$ were reduced, this would favor configurations with ferromagnetic correlations such as the right side of Fig.~\ref{fig:exact_XX}, i.e.~the soft domain wall is increasingly biased towards deforming into hard domain walls. 
		
		Returning to the SUSY limit, the quantum fluctuations of the domain wall can thus be captured by a uniform XX model at zero $S_z$ magnetization. [Note that a non-zero magnetization corresponds to an overall slope in the domain wall.] Via Jordan-Wigner transformation, the kinetic energy is just given by the energy of the half-filled spinless tight-binding chain with hopping parameter $t$. This gives an energy of $\frac{1}{2\pi}\int_{-\frac{\pi}{2}}^{\frac{\pi}{2}}dk(-2|t|\cos k)=\frac{-2|t|}{\pi}$ per spin degree of freedom. To get the contribution per doped hole, we recall that the domain wall accommodates $2/3$ holes worth of charge per unit length in the horizontal direction. Since each unit length has two spins, this leads to a kinetic energy of $\frac{-6|t|}{\pi}$ per hole. 
		
		Finally we need to account for the potential energy difference with the uniform state at $\nu=1/2$. Neglecting the overall constant shift in Eq.~\ref{eq:H_susy_honeycomb}, the energy of a honeycomb plaquette embedded in a sublattice polarized region is $V_\Delta+3V_2+\mu$, while the energy of a plaquette in the domain wall (Fig.~\ref{fig:exact_XX} left) is $\frac{V_\triangle}{3}+V_2+\frac{2\mu}{3}$. In the SUSY limit, the relative energy of a domain wall plaquette is $-\frac{2V_\triangle}{3}-2V_2-\frac{\mu}{3}=-\frac{1}{3}$. However, there are two domain wall plaquettes per unit length, which together contribute $\frac{2}{3}$ holes. Hence the potential energy per hole is $-1$, leading to a domain wall line tension of $\epsilon_{\text{DW}}=-(1+\frac{6}{\pi})$ per hole. 
		
		In the mean-field analysis of Sec.~\ref{subsubsec:MFbubbles}, the boundary A fermions in the left of Fig.~\ref{fig:exact_XX} are only allowed to delocalize vertically. In the spin language, this corresponds to only having an XX interaction between disjoint pairs of adjacent spins, leading to a line tension $\epsilon^{\text{MF}}_{\text{DW}}=-(1+\frac{3}{2})$.
		
		\section{Clusters with zero energy states}
		
		This section contains raw data of degeneracy $N_0$ and excitation gaps $\Delta_{\mathrm{ex}}$ for all simulation clusters that were found to have zero ground state energy in our ED study. Table~\ref{tab:clusters} is sorted first by lowest $\nu=1/4$ to highest filling $\nu=0.312$ (though we believe $\nu=0.312$ is beyond the true upper limit of the zero-energy window, as discussed in the main text) and then from smallest $N=14$ to largest system size $N=56$. In addition to the ED output consisting of total zero degeneracy $N_0$, maximum zero degeneracy within a (charge, translation and point group) symmetry sector $\max_{\alpha}N_0^{\alpha}$ and the excitation gap corresponding to the first non-zero eigenvalue across all translation and point group symmetry sectors $\Delta_{\mathrm{ex}}$, we state each cluster's periodic simulation cell $\mathbf{u}_1,\mathbf{u}_2$, as described in Sec.~\ref{subsec:methods}, as well as its point group (PG) and aspect ratio.
		
		\begin{table*}
			\caption{Overview of simulation clusters with zero-energy states used for ED in this work. 
				\label{tab:clusters}}
			\begin{minipage}[t]{0.49\textwidth}\vspace{0pt}
				\centering
				\begin{tabular}{P{0.7cm} | P{0.4cm} P{1cm} P{1cm} P{0.4cm} P{0.8cm} | P{0.5cm} P{1.1cm} P{0.8cm}}
					\hline
					\hline
					$\nu$ & $N$ & $\mathbf{u}_1$ & $\mathbf{u}_2$ & PG & aspect & $N_0$ & $\max_{\alpha}N_0^{\alpha}$  & $\Delta_{\mathrm{ex}}$ \\ 
					\hline
					0.250 & 16 & $(1,2)$ & $(3,-2)$ & $D_2$ & 1.00 & 4 & 1 & 0.7811 \\ 
					& 20 & $(1,2)$ & $(4,-2)$ & $C_2$ & 1.33 & 1 & 1 & 0.1330 \\ 
					& 24 & $(2,2)$ & $(2,-4)$ & $D_6$ & 1.00 & 4 & 1 & 0.7038 \\ 
					& 24 & $(1,4)$ & $(3,0)$ & $D_2$ & 1.33 & 1 & 1 & 0.3873 \\ 
					& 24 & $(1,2)$ & $(5,-2)$ & $C_2$ & 1.33 & 4 & 1 & 0.2339 \\ 
					& 28 & $(1,3)$ & $(4,-2)$ & $C_2$ & 1.00 & 1 & 1 & 0.1746 \\ 
					& 32 & $(4,0)$ & $(0,4)$ & $D_6$ & 1.00 & 15 & 2 & 0.9914 \\ 
					& 32 & $(2,2)$ & $(4,-4)$ & $D_2$ & 1.00 & 8 & 2 & 0.4658 \\ 
					& 32 & $(2,-4)$ & $(3,2)$ & $D_2$ & 1.25 & 5 & 2 & 0.2411 \\ 
					& 32 & $(1,3)$ & $(5,-1)$ & $C_2$ & 1.00 & 8 & 1 & 0.1932 \\ 
					& 36 & $(1,4)$ & $(4,-2)$ & $C_2$ & 1.25 & 1 & 1 & 0.0420 \\ 
					& 36 & $(1,3)$ & $(5,-3)$ & $C_2$ & 1.25 & 1 & 1 & 0.1392 \\ 
					& 40 & $(1,4)$ & $(5,0)$ & $C_2$ & 1.25 & 1 & 1 & 0.1158 \\ 
					& 40 & $(1,5)$ & $(4,0)$ & $C_2$ & 1.25 & 4 & 1 & 0.2465 \\ 
					& 40 & $(2,-4)$ & $(3,4)$ & $D_2$ & 1.25 & 2 & 1 & 0.1597 \\ 
					& 40 & $(2,2)$ & $(4,-6)$ & $D_2$ & 1.50 & 5 & 2 & 0.2565 \\ 
					& 44 & $(1,5)$ & $(4,-2)$ & $C_2$ & 1.50 & 1 & 1 & 0.0786 \\ 
					& 48 & $(1,4)$ & $(5,-4)$ & $D_2$ & 1.00 & 8 & 1 & 0.3443 \\ 
					& 48 & $(2,2)$ & $(6,-6)$ & $D_2$ & 1.50 & 9 & 3 & 0.2236 \\ 
					& 48 & $(1,6)$ & $(4,0)$ & $D_2$ & 1.50 & 17 & 2 & 0.0869 \\ 
					& 48 & $(1,5)$ & $(5,1)$ & $D_2$ & 1.50 & 16 & 2 & 0.2825 \\ 
					& 48 & $(3,3)$ & $(5,-3)$ & $C_2$ & 1.20 & 4 & 1 & 0.3087 \\ 
					& 48 & $(2,4)$ & $(4,-4)$ & $C_2$ & 1.50 & 20 & 3 & 0.1774 \\ 
					& 48 & $(1,3)$ & $(5,-9)$ & $C_2$ & 1.50 & 8 & 1 & 0.3152 \\ 
					& 56 & $(2,4)$ & $(6,-2)$ & $C_6$ & 1.00 & 4 & 1 & 0.2817 \\ 
					& 56 & $(1,5)$ & $(5,-3)$ & $C_2$ & 1.20 & 4 & 1 & 0.2918 \\ 
					0.259 & 54 & $(3,3)$ & $(3,-6)$ & $D_6$ & 1.00 & 2 & 1 & 0.0915 \\ 
					& 54 & $(1,5)$ & $(5,-2)$ & $C_2$ & 1.20 & 53 & 2 & 0.1403 \\ 
					& 54 & $(1,4)$ & $(5,-7)$ & $C_2$ & 1.20 & 28 & 2 & 0.0118 \\ 
					0.260 & 50 & $(5,0)$ & $(0,5)$ & $D_6$ & 1.00 & 52 & 2 & 0.1061 \\ 
					& 50 & $(3,2)$ & $(5,-5)$ & $C_2$ & 1.00 & 52 & 4 & 0.0522 \\ 
					& 50 & $(1,4)$ & $(5,-5)$ & $C_2$ & 1.00 & 27 & 3 & 0.0242 \\ 
					0.261 & 46 & $(1,4)$ & $(5,-3)$ & $C_2$ & 1.00 & 45 & 2 & 0.0973 \\ 
					0.262 & 42 & $(1,4)$ & $(5,-1)$ & $C_6$ & 1.00 & 36 & 2 & 0.3135 \\ 
					& 42 & $(2,-5)$ & $(3,3)$ & $D_2$ & 1.00 & 43 & 2 & 0.4319 \\ 
					& 42 & $(1,3)$ & $(5,-6)$ & $C_2$ & 1.50 & 1 & 1 & 0.0409 \\ 
					0.263 & 38 & $(2,-5)$ & $(3,2)$ & $C_6$ & 1.00 & 37 & 2 & 0.5755 \\ 
					& 38 & $(1,3)$ & $(5,-4)$ & $C_2$ & 1.25 & 18 & 1 & 0.4848 \\ 
					0.265 & 34 & $(1,3)$ & $(5,-2)$ & $C_2$ & 1.25 & 18 & 2 & 0.6710 \\ 
					0.267 & 30 & $(1,3)$ & $(4,-3)$ & $D_2$ & 1.00 & 2 & 1 & 0.1684 \\ 
					0.268 & 56 & $(1,5)$ & $(5,-3)$ & $C_2$ & 1.20 & 113 & 5 & 0.2918 \\ 
					& 56 & $(1,4)$ & $(7,0)$ & $C_2$ & 1.20 & 78 & 3 & 0.2231 \\ 
				\end{tabular}
			\end{minipage} \hfill
			\begin{minipage}[t]{0.49\textwidth}\vspace{1.5em}
				\centering
				\begin{tabular}{P{0.7cm} | P{0.4cm} P{1cm} P{1cm} P{0.4cm} P{0.8cm} | P{0.5cm} P{1.1cm} P{0.8cm}}
					0.269 & 26 & $(1,3)$ & $(4,-1)$ & $C_6$ & 1.00 & 1 & 1 & 0.6396 \\ 
					& 52 & $(3,-5)$ & $(4,2)$ & $C_2$ & 1.20 & 64 & 3 & 0.0868 \\ 
					& 52 & $(1,4)$ & $(5,-6)$ & $C_2$ & 1.20 & 38 & 2 & 0.1103 \\ 
					0.271 & 48 & $(1,4)$ & $(5,-4)$ & $D_2$ & 1.00 & 49 & 2 & 0.3443 \\ 
					& 48 & $(2,2)$ & $(6,-6)$ & $D_2$ & 1.50 & 2 & 1 & 0.2236 \\ 
					& 48 & $(1,6)$ & $(4,0)$ & $D_2$ & 1.50 & 2 & 1 & 0.0869 \\ 
					& 48 & $(1,5)$ & $(5,1)$ & $D_2$ & 1.50 & 1 & 1 & 0.2825 \\ 
					& 48 & $(3,3)$ & $(5,-3)$ & $C_2$ & 1.20 & 52 & 3 & 0.3087 \\ 
					& 48 & $(2,4)$ & $(4,-4)$ & $C_2$ & 1.50 & 1 & 1 & 0.1774 \\ 
					& 48 & $(1,3)$ & $(5,-9)$ & $C_2$ & 1.50 & 4 & 1 & 0.3152 \\ 
					0.273 & 22 & $(1,2)$ & $(4,-3)$ & $C_2$ & 1.33 & 10 & 1 & 0.8924 \\ 
					& 44 & $(1,4)$ & $(5,-2)$ & $C_2$ & 1.00 & 32 & 2 & 0.0793 \\ 
					& 44 & $(1,3)$ & $(5,-7)$ & $C_2$ & 1.50 & 10 & 1 & 0.1132 \\ 
					0.275 & 40 & $(2,-4)$ & $(3,4)$ & $D_2$ & 1.25 & 3 & 2 & 0.1597 \\ 
					& 40 & $(2,2)$ & $(4,-6)$ & $D_2$ & 1.50 & 2 & 1 & 0.2565 \\ 
					& 40 & $(1,5)$ & $(4,0)$ & $C_2$ & 1.25 & 11 & 2 & 0.2465 \\ 
					& 40 & $(1,4)$ & $(5,0)$ & $C_2$ & 1.25 & 12 & 3 & 0.1158 \\ 
					& 40 & $(1,3)$ & $(5,-5)$ & $C_2$ & 1.25 & 21 & 2 & 0.3259 \\ 
					0.278 & 18 & $(3,0)$ & $(0,3)$ & $D_6$ & 1.00 & 10 & 1 & 1.5401 \\ 
					& 18 & $(1,2)$ & $(4,-1)$ & $C_2$ & 1.00 & 7 & 1 & 0.8392 \\ 
					& 36 & $(1,3)$ & $(5,-3)$ & $C_2$ & 1.25 & 6 & 1 & 0.1392 \\ 
					& 54 & $(3,3)$ & $(3,-6)$ & $D_6$ & 1.00 & 66 & 2 & 0.0915 \\ 
					& 54 & $(1,5)$ & $(5,-2)$ & $C_2$ & 1.20 & 6 & 1 & 0.1403 \\ 
					& 54 & $(1,4)$ & $(5,-7)$ & $C_2$ & 1.20 & 2 & 2 & 0.0118 \\ 
					0.280 & 50 & $(5,0)$ & $(0,5)$ & $D_6$ & 1.00 & 1 & 1 & 0.1061 \\ 
					& 50 & $(3,2)$ & $(5,-5)$ & $C_2$ & 1.00 & 1 & 1 & 0.0522 \\ 
					& 50 & $(1,4)$ & $(5,-5)$ & $C_2$ & 1.00 & 1 & 1 & 0.0242 \\ 
					0.281 & 32 & $(2,2)$ & $(4,-4)$ & $D_2$ & 1.00 & 1 & 1 & 0.4658 \\ 
					& 32 & $(2,-4)$ & $(3,2)$ & $D_2$ & 1.25 & 2 & 1 & 0.2411 \\ 
					& 32 & $(1,3)$ & $(5,-1)$ & $C_2$ & 1.00 & 1 & 1 & 0.1932 \\ 
					0.283 & 46 & $(1,3)$ & $(5,-8)$ & $C_2$ & 1.50 & 1 & 1 & 0.0168 \\ 
					0.286 & 14 & $(1,2)$ & $(3,-1)$ & $C_6$ & 1.00 & 6 & 1 & 1.3944 \\ 
					& 42 & $(2,-5)$ & $(3,3)$ & $D_2$ & 1.00 & 3 & 1 & 0.4319 \\ 
					& 42 & $(1,3)$ & $(5,-6)$ & $C_2$ & 1.50 & 3 & 1 & 0.0409 \\ 
					& 56 & $(2,4)$ & $(6,-2)$ & $C_6$ & 1.00 & 6 & 1 & 0.0577 \\ 
					0.292 & 24 & $(2,2)$ & $(2,-4)$ & $D_6$ & 1.00 & 1 & 1 & 0.7038 \\ 
					& 24 & $(1,4)$ & $(3,0)$ & $D_2$ & 1.33 & 2 & 1 & 0.3873 \\ 
					& 24 & $(1,3)$ & $(4,0)$ & $C_2$ & 1.33 & 4 & 1 & 0.3667 \\ 
					& 24 & $(1,2)$ & $(5,-2)$ & $C_2$ & 1.33 & 4 & 1 & 0.2339 \\ 
					& 48 & $(1,6)$ & $(4,0)$ & $D_2$ & 1.50 & 3 & 1 & 0.3847 \\ 
					& 48 & $(1,5)$ & $(5,1)$ & $D_2$ & 1.50 & 3 & 1 & 0.3576 \\ 
					& 48 & $(2,4)$ & $(4,-4)$ & $C_2$ & 1.50 & 3 & 1 & 0.3256 \\ 
					0.312 & 16 & $(1,2)$ & $(3,-2)$ & $D_2$ & 1.00 & 1 & 1 & 0.7811 \\ 
					\hline
					\hline
				\end{tabular}
			\end{minipage}
		\end{table*}
		
	\end{appendix}
	

\begin{thebibliography}{41}%
		\makeatletter
		\providecommand \@ifxundefined [1]{%
			\@ifx{#1\undefined}
		}%
		\providecommand \@ifnum [1]{%
			\ifnum #1\expandafter \@firstoftwo
			\else \expandafter \@secondoftwo
			\fi
		}%
		\providecommand \@ifx [1]{%
			\ifx #1\expandafter \@firstoftwo
			\else \expandafter \@secondoftwo
			\fi
		}%
		\providecommand \natexlab [1]{#1}%
		\providecommand \enquote  [1]{``#1''}%
		\providecommand \bibnamefont  [1]{#1}%
		\providecommand \bibfnamefont [1]{#1}%
		\providecommand \citenamefont [1]{#1}%
		\providecommand \href@noop [0]{\@secondoftwo}%
		\providecommand \href [0]{\begingroup \@sanitize@url \@href}%
		\providecommand \@href[1]{\@@startlink{#1}\@@href}%
		\providecommand \@@href[1]{\endgroup#1\@@endlink}%
		\providecommand \@sanitize@url [0]{\catcode `\\12\catcode `\$12\catcode
			`\&12\catcode `\#12\catcode `\^12\catcode `\_12\catcode `\%12\relax}%
		\providecommand \@@startlink[1]{}%
		\providecommand \@@endlink[0]{}%
		\providecommand \url  [0]{\begingroup\@sanitize@url \@url }%
		\providecommand \@url [1]{\endgroup\@href {#1}{\urlprefix }}%
		\providecommand \urlprefix  [0]{URL }%
		\providecommand \Eprint [0]{\href }%
		\providecommand \doibase [0]{https://doi.org/}%
		\providecommand \selectlanguage [0]{\@gobble}%
		\providecommand \bibinfo  [0]{\@secondoftwo}%
		\providecommand \bibfield  [0]{\@secondoftwo}%
		\providecommand \translation [1]{[#1]}%
		\providecommand \BibitemOpen [0]{}%
		\providecommand \bibitemStop [0]{}%
		\providecommand \bibitemNoStop [0]{.\EOS\space}%
		\providecommand \EOS [0]{\spacefactor3000\relax}%
		\providecommand \BibitemShut  [1]{\csname bibitem#1\endcsname}%
		\let\auto@bib@innerbib\@empty
		%</preamble>
		\bibitem [{\citenamefont {Cao}\ \emph {et~al.}(2018{\natexlab{a}})\citenamefont
			{Cao}, \citenamefont {Fatemi}, \citenamefont {Demir}, \citenamefont {Fang},
			\citenamefont {Tomarken}, \citenamefont {Luo}, \citenamefont
			{Sanchez-Yamagishi}, \citenamefont {Watanabe}, \citenamefont {Taniguchi},
			\citenamefont {Kaxiras}, \citenamefont {Ashoori},\ and\ \citenamefont
			{Jarillo-Herrero}}]{Cao2018correlated}%
		\BibitemOpen
		\bibfield  {author} {\bibinfo {author} {\bibfnamefont {Y.}~\bibnamefont
				{Cao}}, \bibinfo {author} {\bibfnamefont {V.}~\bibnamefont {Fatemi}},
			\bibinfo {author} {\bibfnamefont {A.}~\bibnamefont {Demir}}, \bibinfo
			{author} {\bibfnamefont {S.}~\bibnamefont {Fang}}, \bibinfo {author}
			{\bibfnamefont {S.~L.}\ \bibnamefont {Tomarken}}, \bibinfo {author}
			{\bibfnamefont {J.~Y.}\ \bibnamefont {Luo}}, \bibinfo {author} {\bibfnamefont
				{J.~D.}\ \bibnamefont {Sanchez-Yamagishi}}, \bibinfo {author} {\bibfnamefont
				{K.}~\bibnamefont {Watanabe}}, \bibinfo {author} {\bibfnamefont
				{T.}~\bibnamefont {Taniguchi}}, \bibinfo {author} {\bibfnamefont
				{E.}~\bibnamefont {Kaxiras}}, \bibinfo {author} {\bibfnamefont {R.~C.}\
				\bibnamefont {Ashoori}},\ and\ \bibinfo {author} {\bibfnamefont
				{P.}~\bibnamefont {Jarillo-Herrero}},\ }\bibfield  {title} {\bibinfo {title}
			{Correlated insulator behaviour at half-filling in magic-angle graphene
				superlattices},\ }\href {https://doi.org/10.1038/nature26154} {\bibfield
			{journal} {\bibinfo  {journal} {Nature}\ }\textbf {\bibinfo {volume} {556}},\
			\bibinfo {pages} {80} (\bibinfo {year} {2018}{\natexlab{a}})}\BibitemShut
		{NoStop}%
		\bibitem [{\citenamefont {Cao}\ \emph {et~al.}(2018{\natexlab{b}})\citenamefont
			{Cao}, \citenamefont {Fatemi}, \citenamefont {Fang}, \citenamefont
			{Watanabe}, \citenamefont {Taniguchi}, \citenamefont {Kaxiras},\ and\
			\citenamefont {Jarillo-Herrero}}]{Cao2018unconventional}%
		\BibitemOpen
		\bibfield  {author} {\bibinfo {author} {\bibfnamefont {Y.}~\bibnamefont
				{Cao}}, \bibinfo {author} {\bibfnamefont {V.}~\bibnamefont {Fatemi}},
			\bibinfo {author} {\bibfnamefont {S.}~\bibnamefont {Fang}}, \bibinfo {author}
			{\bibfnamefont {K.}~\bibnamefont {Watanabe}}, \bibinfo {author}
			{\bibfnamefont {T.}~\bibnamefont {Taniguchi}}, \bibinfo {author}
			{\bibfnamefont {E.}~\bibnamefont {Kaxiras}},\ and\ \bibinfo {author}
			{\bibfnamefont {P.}~\bibnamefont {Jarillo-Herrero}},\ }\bibfield  {title}
		{\bibinfo {title} {Unconventional superconductivity in magic-angle graphene
				superlattices},\ }\href {https://doi.org/10.1038/nature26160} {\bibfield
			{journal} {\bibinfo  {journal} {Nature}\ }\textbf {\bibinfo {volume} {556}},\
			\bibinfo {pages} {43} (\bibinfo {year} {2018}{\natexlab{b}})}\BibitemShut
		{NoStop}%
		\bibitem [{\citenamefont {Yankowitz}\ \emph {et~al.}(2019)\citenamefont
			{Yankowitz}, \citenamefont {Chen}, \citenamefont {Polshyn}, \citenamefont
			{Zhang}, \citenamefont {Watanabe}, \citenamefont {Taniguchi}, \citenamefont
			{Graf}, \citenamefont {Young},\ and\ \citenamefont
			{Dean}}]{Yankowitz2019tuning}%
		\BibitemOpen
		\bibfield  {author} {\bibinfo {author} {\bibfnamefont {M.}~\bibnamefont
				{Yankowitz}}, \bibinfo {author} {\bibfnamefont {S.}~\bibnamefont {Chen}},
			\bibinfo {author} {\bibfnamefont {H.}~\bibnamefont {Polshyn}}, \bibinfo
			{author} {\bibfnamefont {Y.}~\bibnamefont {Zhang}}, \bibinfo {author}
			{\bibfnamefont {K.}~\bibnamefont {Watanabe}}, \bibinfo {author}
			{\bibfnamefont {T.}~\bibnamefont {Taniguchi}}, \bibinfo {author}
			{\bibfnamefont {D.}~\bibnamefont {Graf}}, \bibinfo {author} {\bibfnamefont
				{A.~F.}\ \bibnamefont {Young}},\ and\ \bibinfo {author} {\bibfnamefont
				{C.~R.}\ \bibnamefont {Dean}},\ }\bibfield  {title} {\bibinfo {title} {Tuning
				superconductivity in twisted bilayer graphene},\ }\href
		{https://doi.org/10.1126/science.aav1910} {\bibfield  {journal} {\bibinfo
				{journal} {Science}\ }\textbf {\bibinfo {volume} {363}},\ \bibinfo {pages}
			{1059} (\bibinfo {year} {2019})}\BibitemShut {NoStop}%
		\bibitem [{\citenamefont {Lu}\ \emph {et~al.}(2019)\citenamefont {Lu},
			\citenamefont {Stepanov}, \citenamefont {Yang}, \citenamefont {Xie},
			\citenamefont {Aamir}, \citenamefont {Das}, \citenamefont {Urgell},
			\citenamefont {Watanabe}, \citenamefont {Taniguchi}, \citenamefont {Zhang},
			\citenamefont {Bachtold}, \citenamefont {MacDonald},\ and\ \citenamefont
			{Efetov}}]{Lu2019orbital}%
		\BibitemOpen
		\bibfield  {author} {\bibinfo {author} {\bibfnamefont {X.}~\bibnamefont
				{Lu}}, \bibinfo {author} {\bibfnamefont {P.}~\bibnamefont {Stepanov}},
			\bibinfo {author} {\bibfnamefont {W.}~\bibnamefont {Yang}}, \bibinfo {author}
			{\bibfnamefont {M.}~\bibnamefont {Xie}}, \bibinfo {author} {\bibfnamefont
				{M.~A.}\ \bibnamefont {Aamir}}, \bibinfo {author} {\bibfnamefont
				{I.}~\bibnamefont {Das}}, \bibinfo {author} {\bibfnamefont {C.}~\bibnamefont
				{Urgell}}, \bibinfo {author} {\bibfnamefont {K.}~\bibnamefont {Watanabe}},
			\bibinfo {author} {\bibfnamefont {T.}~\bibnamefont {Taniguchi}}, \bibinfo
			{author} {\bibfnamefont {G.}~\bibnamefont {Zhang}}, \bibinfo {author}
			{\bibfnamefont {A.}~\bibnamefont {Bachtold}}, \bibinfo {author}
			{\bibfnamefont {A.~H.}\ \bibnamefont {MacDonald}},\ and\ \bibinfo {author}
			{\bibfnamefont {D.~K.}\ \bibnamefont {Efetov}},\ }\bibfield  {title}
		{\bibinfo {title} {Superconductors, orbital magnets and correlated states in
				magic-angle bilayer graphene},\ }\href
		{https://doi.org/10.1038/s41586-019-1695-0} {\bibfield  {journal} {\bibinfo
				{journal} {Nature}\ }\textbf {\bibinfo {volume} {574}},\ \bibinfo {pages}
			{653} (\bibinfo {year} {2019})}\BibitemShut {NoStop}%
		\bibitem [{\citenamefont {Koshino}\ \emph {et~al.}(2018)\citenamefont
			{Koshino}, \citenamefont {Yuan}, \citenamefont {Koretsune}, \citenamefont
			{Ochi}, \citenamefont {Kuroki},\ and\ \citenamefont
			{Fu}}]{Koshino2018maximally}%
		\BibitemOpen
		\bibfield  {author} {\bibinfo {author} {\bibfnamefont {M.}~\bibnamefont
				{Koshino}}, \bibinfo {author} {\bibfnamefont {N.~F.~Q.}\ \bibnamefont
				{Yuan}}, \bibinfo {author} {\bibfnamefont {T.}~\bibnamefont {Koretsune}},
			\bibinfo {author} {\bibfnamefont {M.}~\bibnamefont {Ochi}}, \bibinfo {author}
			{\bibfnamefont {K.}~\bibnamefont {Kuroki}},\ and\ \bibinfo {author}
			{\bibfnamefont {L.}~\bibnamefont {Fu}},\ }\bibfield  {title} {\bibinfo
			{title} {{Maximally Localized Wannier Orbitals and the Extended Hubbard Model
					for Twisted Bilayer Graphene}},\ }\href
		{https://doi.org/10.1103/PhysRevX.8.031087} {\bibfield  {journal} {\bibinfo
				{journal} {Phys. Rev. X}\ }\textbf {\bibinfo {volume} {8}},\ \bibinfo {pages}
			{031087} (\bibinfo {year} {2018})}\BibitemShut {NoStop}%
		\bibitem [{\citenamefont {Kang}\ and\ \citenamefont
			{Vafek}(2018)}]{Kang2018symmetry}%
		\BibitemOpen
		\bibfield  {author} {\bibinfo {author} {\bibfnamefont {J.}~\bibnamefont
				{Kang}}\ and\ \bibinfo {author} {\bibfnamefont {O.}~\bibnamefont {Vafek}},\
		}\bibfield  {title} {\bibinfo {title} {{Symmetry, Maximally Localized Wannier
					States, and a Low-Energy Model for Twisted Bilayer Graphene Narrow Bands}},\
		}\href {https://doi.org/10.1103/PhysRevX.8.031088} {\bibfield  {journal}
			{\bibinfo  {journal} {Phys. Rev. X}\ }\textbf {\bibinfo {volume} {8}},\
			\bibinfo {pages} {031088} (\bibinfo {year} {2018})}\BibitemShut {NoStop}%
		\bibitem [{\citenamefont {Zou}\ \emph {et~al.}(2018)\citenamefont {Zou},
			\citenamefont {Po}, \citenamefont {Vishwanath},\ and\ \citenamefont
			{Senthil}}]{Zou2018band}%
		\BibitemOpen
		\bibfield  {author} {\bibinfo {author} {\bibfnamefont {L.}~\bibnamefont
				{Zou}}, \bibinfo {author} {\bibfnamefont {H.~C.}\ \bibnamefont {Po}},
			\bibinfo {author} {\bibfnamefont {A.}~\bibnamefont {Vishwanath}},\ and\
			\bibinfo {author} {\bibfnamefont {T.}~\bibnamefont {Senthil}},\ }\bibfield
		{title} {\bibinfo {title} {{Band structure of twisted bilayer graphene:
					Emergent symmetries, commensurate approximants, and Wannier obstructions}},\
		}\href {https://doi.org/10.1103/PhysRevB.98.085435} {\bibfield  {journal}
			{\bibinfo  {journal} {Phys. Rev. B}\ }\textbf {\bibinfo {volume} {98}},\
			\bibinfo {pages} {085435} (\bibinfo {year} {2018})}\BibitemShut {NoStop}%
		\bibitem [{\citenamefont {Zhang}\ \emph {et~al.}(2022)\citenamefont {Zhang},
			\citenamefont {Zhang}, \citenamefont {Fu},\ and\ \citenamefont
			{Kim}}]{Zhang2022fractional}%
		\BibitemOpen
		\bibfield  {author} {\bibinfo {author} {\bibfnamefont {K.}~\bibnamefont
				{Zhang}}, \bibinfo {author} {\bibfnamefont {Y.}~\bibnamefont {Zhang}},
			\bibinfo {author} {\bibfnamefont {L.}~\bibnamefont {Fu}},\ and\ \bibinfo
			{author} {\bibfnamefont {E.-A.}\ \bibnamefont {Kim}},\ }\bibfield  {title}
		{\bibinfo {title} {{Fractional correlated insulating states at one-third
					filled magic angle twisted bilayer graphene}},\ }\href
		{https://doi.org/10.1038/s42005-022-01027-6} {\bibfield  {journal} {\bibinfo
				{journal} {Commun. Phys.}\ }\textbf {\bibinfo {volume} {5}},\ \bibinfo
			{pages} {1} (\bibinfo {year} {2022})}\BibitemShut {NoStop}%
		\bibitem [{\citenamefont {Mao}\ \emph {et~al.}(2022)\citenamefont {Mao},
			\citenamefont {Zhang},\ and\ \citenamefont {Kim}}]{Mao2022fractionalization}%
		\BibitemOpen
		\bibfield  {author} {\bibinfo {author} {\bibfnamefont {D.}~\bibnamefont
				{Mao}}, \bibinfo {author} {\bibfnamefont {K.}~\bibnamefont {Zhang}},\ and\
			\bibinfo {author} {\bibfnamefont {E.-A.}\ \bibnamefont {Kim}},\ }\href@noop
		{} {\bibinfo {title} {Fractionalization in fractional correlated insulating
				states at $n\pm 1/3$ filled twisted bilayer graphene}} (\bibinfo {year}
		{2022}),\ \Eprint {https://arxiv.org/abs/2211.11622} {arXiv:2211.11622
			[cond-mat.str-el]} \BibitemShut {NoStop}%
		\bibitem [{\citenamefont {Fendley}\ \emph
			{et~al.}(2003{\natexlab{a}})\citenamefont {Fendley}, \citenamefont
			{Schoutens},\ and\ \citenamefont {de~Boer}}]{Fendley2003lattice}%
		\BibitemOpen
		\bibfield  {author} {\bibinfo {author} {\bibfnamefont {P.}~\bibnamefont
				{Fendley}}, \bibinfo {author} {\bibfnamefont {K.}~\bibnamefont {Schoutens}},\
			and\ \bibinfo {author} {\bibfnamefont {J.}~\bibnamefont {de~Boer}},\
		}\bibfield  {title} {\bibinfo {title} {{Lattice Models with $\mathcal{N}=2$
					Supersymmetry}},\ }\href {https://doi.org/10.1103/PhysRevLett.90.120402}
		{\bibfield  {journal} {\bibinfo  {journal} {Phys. Rev. Lett.}\ }\textbf
			{\bibinfo {volume} {90}},\ \bibinfo {pages} {120402} (\bibinfo {year}
			{2003}{\natexlab{a}})}\BibitemShut {NoStop}%
		\bibitem [{\citenamefont {Fendley}\ \emph
			{et~al.}(2003{\natexlab{b}})\citenamefont {Fendley}, \citenamefont
			{Nienhuis},\ and\ \citenamefont {Schoutens}}]{Fendley2003bethe}%
		\BibitemOpen
		\bibfield  {author} {\bibinfo {author} {\bibfnamefont {P.}~\bibnamefont
				{Fendley}}, \bibinfo {author} {\bibfnamefont {B.}~\bibnamefont {Nienhuis}},\
			and\ \bibinfo {author} {\bibfnamefont {K.}~\bibnamefont {Schoutens}},\
		}\bibfield  {title} {\bibinfo {title} {Lattice fermion models with
				supersymmetry},\ }\href {https://doi.org/10.1088/0305-4470/36/50/004}
		{\bibfield  {journal} {\bibinfo  {journal} {Journal of Physics A:
					Mathematical and General}\ }\textbf {\bibinfo {volume} {36}},\ \bibinfo
			{pages} {12399} (\bibinfo {year} {2003}{\natexlab{b}})}\BibitemShut {NoStop}%
		\bibitem [{\citenamefont {Fendley}\ and\ \citenamefont
			{Schoutens}(2005)}]{fendley2005susy}%
		\BibitemOpen
		\bibfield  {author} {\bibinfo {author} {\bibfnamefont {P.}~\bibnamefont
				{Fendley}}\ and\ \bibinfo {author} {\bibfnamefont {K.}~\bibnamefont
				{Schoutens}},\ }\bibfield  {title} {\bibinfo {title} {Exact results for
				strongly correlated fermions in $2+1$ dimensions},\ }\href
		{https://doi.org/10.1103/PhysRevLett.95.046403} {\bibfield  {journal}
			{\bibinfo  {journal} {Phys. Rev. Lett.}\ }\textbf {\bibinfo {volume} {95}},\
			\bibinfo {pages} {046403} (\bibinfo {year} {2005})}\BibitemShut {NoStop}%
		\bibitem [{\citenamefont {Huijse}\ and\ \citenamefont
			{Schoutens}(2008)}]{huijse2008review}%
		\BibitemOpen
		\bibfield  {author} {\bibinfo {author} {\bibfnamefont {L.}~\bibnamefont
				{Huijse}}\ and\ \bibinfo {author} {\bibfnamefont {K.}~\bibnamefont
				{Schoutens}},\ }\bibfield  {title} {\bibinfo {title} {Superfrustration of
				charge degrees of freedom},\ }\href@noop {} {\bibfield  {journal} {\bibinfo
				{journal} {The European Physical Journal B}\ }\textbf {\bibinfo {volume}
				{64}},\ \bibinfo {pages} {543} (\bibinfo {year} {2008})}\BibitemShut
		{NoStop}%
		\bibitem [{\citenamefont {Witten}(1982)}]{Witten1982constraints}%
		\BibitemOpen
		\bibfield  {author} {\bibinfo {author} {\bibfnamefont {E.}~\bibnamefont
				{Witten}},\ }\bibfield  {title} {\bibinfo {title} {Constraints on
				supersymmetry breaking},\ }\href
		{https://doi.org/10.1016/0550-3213(82)90071-2} {\bibfield  {journal}
			{\bibinfo  {journal} {Nuclear Physics B}\ }\textbf {\bibinfo {volume}
				{202}},\ \bibinfo {pages} {253} (\bibinfo {year} {1982})}\BibitemShut
		{NoStop}%
		\bibitem [{\citenamefont {van Eerten}(2005)}]{vaneerten2005witten}%
		\BibitemOpen
		\bibfield  {author} {\bibinfo {author} {\bibfnamefont {H.}~\bibnamefont {van
					Eerten}},\ }\bibfield  {title} {\bibinfo {title} {Extensive ground state
				entropy in supersymmetric lattice models},\ }\href
		{https://doi.org/10.1063/1.2142836} {\bibfield  {journal} {\bibinfo
				{journal} {Journal of Mathematical Physics}\ }\textbf {\bibinfo {volume}
				{46}},\ \bibinfo {pages} {123302} (\bibinfo {year} {2005})}\BibitemShut {NoStop}%
		\bibitem [{\citenamefont {Jonsson}(2010)}]{jonsson2010homology}%
		\BibitemOpen
		\bibfield  {author} {\bibinfo {author} {\bibfnamefont {J.}~\bibnamefont
				{Jonsson}},\ }\bibfield  {title} {\bibinfo {title} {Certain homology cycles
				of the independence complex of grids},\ }\href@noop {} {\bibfield  {journal}
			{\bibinfo  {journal} {Discrete \& Computational Geometry}\ }\textbf {\bibinfo
				{volume} {43}},\ \bibinfo {pages} {927} (\bibinfo {year} {2010})}\BibitemShut
		{NoStop}%
		\bibitem [{\citenamefont {Huijse}\ \emph {et~al.}(2008)\citenamefont {Huijse},
			\citenamefont {Halverson}, \citenamefont {Fendley},\ and\ \citenamefont
			{Schoutens}}]{huijse2008susy}%
		\BibitemOpen
		\bibfield  {author} {\bibinfo {author} {\bibfnamefont {L.}~\bibnamefont
				{Huijse}}, \bibinfo {author} {\bibfnamefont {J.}~\bibnamefont {Halverson}},
			\bibinfo {author} {\bibfnamefont {P.}~\bibnamefont {Fendley}},\ and\ \bibinfo
			{author} {\bibfnamefont {K.}~\bibnamefont {Schoutens}},\ }\bibfield  {title}
		{\bibinfo {title} {Charge frustration and quantum criticality for strongly
				correlated fermions},\ }\href
		{https://doi.org/10.1103/PhysRevLett.101.146406} {\bibfield  {journal}
			{\bibinfo  {journal} {Phys. Rev. Lett.}\ }\textbf {\bibinfo {volume} {101}},\
			\bibinfo {pages} {146406} (\bibinfo {year} {2008})}\BibitemShut {NoStop}%
		\bibitem [{\citenamefont {Huijse}\ \emph {et~al.}(2012)\citenamefont {Huijse},
			\citenamefont {Mehta}, \citenamefont {Moran}, \citenamefont {Schoutens},\
			and\ \citenamefont {Vala}}]{Huijse2012triangular}%
		\BibitemOpen
		\bibfield  {author} {\bibinfo {author} {\bibfnamefont {L.}~\bibnamefont
				{Huijse}}, \bibinfo {author} {\bibfnamefont {D.}~\bibnamefont {Mehta}},
			\bibinfo {author} {\bibfnamefont {N.}~\bibnamefont {Moran}}, \bibinfo
			{author} {\bibfnamefont {K.}~\bibnamefont {Schoutens}},\ and\ \bibinfo
			{author} {\bibfnamefont {J.}~\bibnamefont {Vala}},\ }\bibfield  {title}
		{\bibinfo {title} {Supersymmetric lattice fermions on the triangular lattice:
				superfrustration and criticality},\ }\href
		{https://doi.org/10.1088/1367-2630/14/7/073002} {\bibfield  {journal}
			{\bibinfo  {journal} {New Journal of Physics}\ }\textbf {\bibinfo {volume}
				{14}},\ \bibinfo {pages} {073002} (\bibinfo {year} {2012})}\BibitemShut
		{NoStop}%
		\bibitem [{\citenamefont {Huijse}\ \emph {et~al.}(2011)\citenamefont {Huijse},
			\citenamefont {Moran}, \citenamefont {Vala},\ and\ \citenamefont
			{Schoutens}}]{Huijse2011staggered}%
		\BibitemOpen
		\bibfield  {author} {\bibinfo {author} {\bibfnamefont {L.}~\bibnamefont
				{Huijse}}, \bibinfo {author} {\bibfnamefont {N.}~\bibnamefont {Moran}},
			\bibinfo {author} {\bibfnamefont {J.}~\bibnamefont {Vala}},\ and\ \bibinfo
			{author} {\bibfnamefont {K.}~\bibnamefont {Schoutens}},\ }\bibfield  {title}
		{\bibinfo {title} {{Exact ground states of a staggered supersymmetric model
					for lattice fermions}},\ }\href {https://doi.org/10.1103/PhysRevB.84.115124}
		{\bibfield  {journal} {\bibinfo  {journal} {Phys. Rev. B}\ }\textbf {\bibinfo
				{volume} {84}},\ \bibinfo {pages} {115124} (\bibinfo {year}
			{2011})}\BibitemShut {NoStop}%
		\bibitem [{\citenamefont {Bauer}\ \emph {et~al.}(2013)\citenamefont {Bauer},
			\citenamefont {Huijse}, \citenamefont {Berg}, \citenamefont {Troyer},\ and\
			\citenamefont {Schoutens}}]{Bauer2013multicritical}%
		\BibitemOpen
		\bibfield  {author} {\bibinfo {author} {\bibfnamefont {B.}~\bibnamefont
				{Bauer}}, \bibinfo {author} {\bibfnamefont {L.}~\bibnamefont {Huijse}},
			\bibinfo {author} {\bibfnamefont {E.}~\bibnamefont {Berg}}, \bibinfo {author}
			{\bibfnamefont {M.}~\bibnamefont {Troyer}},\ and\ \bibinfo {author}
			{\bibfnamefont {K.}~\bibnamefont {Schoutens}},\ }\bibfield  {title} {\bibinfo
			{title} {{Supersymmetric multicritical point in a model of lattice
					fermions}},\ }\href {https://doi.org/10.1103/PhysRevB.87.165145} {\bibfield
			{journal} {\bibinfo  {journal} {Phys. Rev. B}\ }\textbf {\bibinfo {volume}
				{87}},\ \bibinfo {pages} {165145} (\bibinfo {year} {2013})}\BibitemShut
		{NoStop}%
		\bibitem [{\citenamefont {Chepiga}\ \emph {et~al.}(2021)\citenamefont
			{Chepiga}, \citenamefont {Minář},\ and\ \citenamefont
			{Schoutens}}]{chepiga2021ladder}%
		\BibitemOpen
		\bibfield  {author} {\bibinfo {author} {\bibfnamefont {N.}~\bibnamefont
				{Chepiga}}, \bibinfo {author} {\bibfnamefont {J.}~\bibnamefont {Minář}},\
			and\ \bibinfo {author} {\bibfnamefont {K.}~\bibnamefont {Schoutens}},\
		}\bibfield  {title} {\bibinfo {title} {{Supersymmetry and multicriticality in
					a ladder of constrained fermions}},\ }\href
		{https://doi.org/10.21468/SciPostPhys.11.3.059} {\bibfield  {journal}
			{\bibinfo  {journal} {SciPost Phys.}\ }\textbf {\bibinfo {volume} {11}},\
			\bibinfo {pages} {59} (\bibinfo {year} {2021})}\BibitemShut {NoStop}%
		\bibitem [{\citenamefont {Galanakis}\ \emph {et~al.}(2012)\citenamefont
			{Galanakis}, \citenamefont {Henley},\ and\ \citenamefont
			{Papanikolaou}}]{galanakis2012triangular}%
		\BibitemOpen
		\bibfield  {author} {\bibinfo {author} {\bibfnamefont {D.}~\bibnamefont
				{Galanakis}}, \bibinfo {author} {\bibfnamefont {C.~L.}\ \bibnamefont
				{Henley}},\ and\ \bibinfo {author} {\bibfnamefont {S.}~\bibnamefont
				{Papanikolaou}},\ }\bibfield  {title} {\bibinfo {title} {Order and
				supersymmetry at high filling zero-energy states on the triangular lattice},\
		}\href {https://doi.org/10.1103/PhysRevB.86.195105} {\bibfield  {journal}
			{\bibinfo  {journal} {Phys. Rev. B}\ }\textbf {\bibinfo {volume} {86}},\
			\bibinfo {pages} {195105} (\bibinfo {year} {2012})}\BibitemShut {NoStop}%
		\bibitem [{\citenamefont {Sarkar}(1991{\natexlab{a}})}]{Sarkar1991tJ}%
		\BibitemOpen
		\bibfield  {author} {\bibinfo {author} {\bibfnamefont {S.}~\bibnamefont
				{Sarkar}},\ }\bibfield  {title} {\bibinfo {title} {The supersymmetric t-j
				model in one dimension},\ }\href {https://doi.org/10.1088/0305-4470/24/5/026}
		{\bibfield  {journal} {\bibinfo  {journal} {Journal of Physics A:
					Mathematical and General}\ }\textbf {\bibinfo {volume} {24}},\ \bibinfo
			{pages} {1137} (\bibinfo {year} {1991}{\natexlab{a}})}\BibitemShut {NoStop}%
		\bibitem [{\citenamefont {Essler}\ and\ \citenamefont
			{Korepin}(1992)}]{Essler1992tJ}%
		\BibitemOpen
		\bibfield  {author} {\bibinfo {author} {\bibfnamefont {F.~H.~L.}\
				\bibnamefont {Essler}}\ and\ \bibinfo {author} {\bibfnamefont {V.~E.}\
				\bibnamefont {Korepin}},\ }\bibfield  {title} {\bibinfo {title} {Higher
				conservation laws and algebraic bethe ansatze for the supersymmetric t-j
				model},\ }\href {https://doi.org/10.1103/physrevb.46.9147} {\bibfield
			{journal} {\bibinfo  {journal} {Physical Review B}\ }\textbf {\bibinfo
				{volume} {46}},\ \bibinfo {pages} {9147} (\bibinfo {year}
			{1992})}\BibitemShut {NoStop}%
		\bibitem [{\citenamefont
			{Sarkar}(1991{\natexlab{b}})}]{Sarkar1991supercoherent}%
		\BibitemOpen
		\bibfield  {author} {\bibinfo {author} {\bibfnamefont {S.}~\bibnamefont
				{Sarkar}},\ }\bibfield  {title} {\bibinfo {title} {Supercoherent states for
				the t-j model},\ }\href {https://doi.org/10.1088/0305-4470/24/24/013}
		{\bibfield  {journal} {\bibinfo  {journal} {Journal of Physics A:
					Mathematical and General}\ }\textbf {\bibinfo {volume} {24}},\ \bibinfo
			{pages} {5775} (\bibinfo {year} {1991}{\natexlab{b}})}\BibitemShut {NoStop}%
		\bibitem [{\citenamefont {Fendley}\ \emph
			{et~al.}(2003{\natexlab{c}})\citenamefont {Fendley}, \citenamefont
			{Nienhuis},\ and\ \citenamefont {Schoutens}}]{Fendley2003SUSY}%
		\BibitemOpen
		\bibfield  {author} {\bibinfo {author} {\bibfnamefont {P.}~\bibnamefont
				{Fendley}}, \bibinfo {author} {\bibfnamefont {B.}~\bibnamefont {Nienhuis}},\
			and\ \bibinfo {author} {\bibfnamefont {K.}~\bibnamefont {Schoutens}},\
		}\bibfield  {title} {\bibinfo {title} {{Lattice fermion models with
					supersymmetry}},\ }\href {https://doi.org/10.1088/0305-4470/36/50/004}
		{\bibfield  {journal} {\bibinfo  {journal} {J. Phys. A: Math. Gen.}\ }\textbf
			{\bibinfo {volume} {36}},\ \bibinfo {pages} {12399} (\bibinfo {year}
			{2003}{\natexlab{c}})}\BibitemShut {NoStop}%
		\bibitem [{\citenamefont {Wietek}\ \emph {et~al.}(2017)\citenamefont {Wietek},
			\citenamefont {Schuler},\ and\ \citenamefont {Läuchli}}]{wietek2017tos}%
		\BibitemOpen
		\bibfield  {author} {\bibinfo {author} {\bibfnamefont {A.}~\bibnamefont
				{Wietek}}, \bibinfo {author} {\bibfnamefont {M.}~\bibnamefont {Schuler}},\
			and\ \bibinfo {author} {\bibfnamefont {A.~M.}\ \bibnamefont {Läuchli}},\
		}\href@noop {} {\bibinfo {title} {Studying continuous symmetry breaking using
				energy level spectroscopy}} (\bibinfo {year} {2017}),\ \Eprint
		{https://arxiv.org/abs/1704.08622} {arXiv:1704.08622 [cond-mat.str-el]}
		\BibitemShut {NoStop}%
		\bibitem [{\citenamefont {Lehoucq}\ \emph {et~al.}(1998)\citenamefont
			{Lehoucq}, \citenamefont {Sorensen},\ and\ \citenamefont
			{Yang}}]{Lehoucq1998arpack}%
		\BibitemOpen
		\bibfield  {author} {\bibinfo {author} {\bibfnamefont {R.~B.}\ \bibnamefont
				{Lehoucq}}, \bibinfo {author} {\bibfnamefont {D.~C.}\ \bibnamefont
				{Sorensen}},\ and\ \bibinfo {author} {\bibfnamefont {C.}~\bibnamefont
				{Yang}},\ }\href@noop {} {\emph {\bibinfo {title} {{ARPACK Users' Guide:
						Solution of Large-scale Eigenvalue Problems with Implicitly Restarted Arnoldi
						Methods}}}}\ (\bibinfo  {publisher} {SIAM},\ \bibinfo {year}
		{1998})\BibitemShut {NoStop}%
		\bibitem [{\citenamefont {Cancs}\ and\ \citenamefont {Bris}(2000)}]{Cancs2000}%
		\BibitemOpen
		\bibfield  {author} {\bibinfo {author} {\bibfnamefont {E.}~\bibnamefont
				{Cancs}}\ and\ \bibinfo {author} {\bibfnamefont {C.~L.}\ \bibnamefont
				{Bris}},\ }\bibfield  {title} {\bibinfo {title} {Can we outperform the {DIIS}
				approach for electronic structure calculations?},\ }\href
		{https://doi.org/10.1002/1097-461x(2000)79:2<82::aid-qua3>3.0.co;2-i}
		{\bibfield  {journal} {\bibinfo  {journal} {International Journal of Quantum
					Chemistry}\ }\textbf {\bibinfo {volume} {79}},\ \bibinfo {pages} {82}
			(\bibinfo {year} {2000})}\BibitemShut {NoStop}%
		\bibitem [{\citenamefont {Kwan}\ \emph {et~al.}(2023)\citenamefont {Kwan},
			\citenamefont {Wilhelm}, \citenamefont {Biswas},\ and\ \citenamefont
			{Parameswaran}}]{Kwan2023hardcore}%
		\BibitemOpen
		\bibfield  {author} {\bibinfo {author} {\bibfnamefont {Y.~H.}\ \bibnamefont
				{Kwan}}, \bibinfo {author} {\bibfnamefont {P.~H.}\ \bibnamefont {Wilhelm}},
			\bibinfo {author} {\bibfnamefont {S.}~\bibnamefont {Biswas}},\ and\ \bibinfo
			{author} {\bibfnamefont {S.~A.}\ \bibnamefont {Parameswaran}},\ }\href@noop
		{} {\bibinfo {title} {Minimal hubbard models of maximal hilbert space
				fragmentation}} (\bibinfo {year} {2023}),\ \Eprint
		{https://arxiv.org/abs/2304.02669} {arXiv:2304.02669 [cond-mat.stat-mech]}
		\BibitemShut {NoStop}%
		\bibitem [{\citenamefont {Mila}(1994)}]{Mila1994}%
		\BibitemOpen
		\bibfield  {author} {\bibinfo {author} {\bibfnamefont {F.}~\bibnamefont
				{Mila}},\ }\bibfield  {title} {\bibinfo {title} {Exact result on the mott
				transition in a two-dimensional model of strongly correlated electrons},\
		}\href {https://doi.org/10.1103/physrevb.49.14047} {\bibfield  {journal}
			{\bibinfo  {journal} {Physical Review B}\ }\textbf {\bibinfo {volume} {49}},\
			\bibinfo {pages} {14047} (\bibinfo {year} {1994})}\BibitemShut {NoStop}%
		\bibitem [{\citenamefont {Henley}\ and\ \citenamefont
			{Zhang}(2001)}]{Henley2001stripes}%
		\BibitemOpen
		\bibfield  {author} {\bibinfo {author} {\bibfnamefont {C.~L.}\ \bibnamefont
				{Henley}}\ and\ \bibinfo {author} {\bibfnamefont {N.-G.}\ \bibnamefont
				{Zhang}},\ }\bibfield  {title} {\bibinfo {title} {{Spinless fermions and
					charged stripes at the strong-coupling limit}},\ }\href
		{https://doi.org/10.1103/PhysRevB.63.233107} {\bibfield  {journal} {\bibinfo
				{journal} {Phys. Rev. B}\ }\textbf {\bibinfo {volume} {63}},\ \bibinfo
			{pages} {233107} (\bibinfo {year} {2001})}\BibitemShut {NoStop}%
		\bibitem [{\citenamefont {Zhang}\ and\ \citenamefont
			{Henley}(2003)}]{Zhang2003stripes}%
		\BibitemOpen
		\bibfield  {author} {\bibinfo {author} {\bibfnamefont {N.~G.}\ \bibnamefont
				{Zhang}}\ and\ \bibinfo {author} {\bibfnamefont {C.~L.}\ \bibnamefont
				{Henley}},\ }\bibfield  {title} {\bibinfo {title} {{Stripes and holes in a
					two-dimensional model of spinless fermions or hardcore bosons}},\ }\href
		{https://doi.org/10.1103/PhysRevB.68.014506} {\bibfield  {journal} {\bibinfo
				{journal} {Phys. Rev. B}\ }\textbf {\bibinfo {volume} {68}},\ \bibinfo
			{pages} {014506} (\bibinfo {year} {2003})}\BibitemShut {NoStop}%
		\bibitem [{\citenamefont {Zhang}\ and\ \citenamefont
			{Henley}(2004)}]{Zhang2004dilute}%
		\BibitemOpen
		\bibfield  {author} {\bibinfo {author} {\bibfnamefont {N.~G.}\ \bibnamefont
				{Zhang}}\ and\ \bibinfo {author} {\bibfnamefont {C.~L.}\ \bibnamefont
				{Henley}},\ }\bibfield  {title} {\bibinfo {title} {Dilute limit of a
				strongly-interacting model of spinless fermions and hardcore bosons on the
				square lattice},\ }\href {https://doi.org/10.1140/epjb/e2004-00135-8}
		{\bibfield  {journal} {\bibinfo  {journal} {The European Physical Journal B}\
			}\textbf {\bibinfo {volume} {38}},\ \bibinfo {pages} {409} (\bibinfo {year}
			{2004})}\BibitemShut {NoStop}%
		\bibitem [{\citenamefont {Henkel}\ \emph {et~al.}(2010)\citenamefont {Henkel},
			\citenamefont {Nath},\ and\ \citenamefont {Pohl}}]{Henkel2010rydberg}%
		\BibitemOpen
		\bibfield  {author} {\bibinfo {author} {\bibfnamefont {N.}~\bibnamefont
				{Henkel}}, \bibinfo {author} {\bibfnamefont {R.}~\bibnamefont {Nath}},\ and\
			\bibinfo {author} {\bibfnamefont {T.}~\bibnamefont {Pohl}},\ }\bibfield
		{title} {\bibinfo {title} {{Three-Dimensional Roton Excitations and
					Supersolid Formation in Rydberg-Excited Bose-Einstein Condensates}},\ }\href
		{https://doi.org/10.1103/PhysRevLett.104.195302} {\bibfield  {journal}
			{\bibinfo  {journal} {Phys. Rev. Lett.}\ }\textbf {\bibinfo {volume} {104}},\
			\bibinfo {pages} {195302} (\bibinfo {year} {2010})}\BibitemShut {NoStop}%
		\bibitem [{\citenamefont {Pupillo}\ \emph {et~al.}(2010)\citenamefont
			{Pupillo}, \citenamefont {Micheli}, \citenamefont {Boninsegni}, \citenamefont
			{Lesanovsky},\ and\ \citenamefont {Zoller}}]{Pupillo2010rydberg}%
		\BibitemOpen
		\bibfield  {author} {\bibinfo {author} {\bibfnamefont {G.}~\bibnamefont
				{Pupillo}}, \bibinfo {author} {\bibfnamefont {A.}~\bibnamefont {Micheli}},
			\bibinfo {author} {\bibfnamefont {M.}~\bibnamefont {Boninsegni}}, \bibinfo
			{author} {\bibfnamefont {I.}~\bibnamefont {Lesanovsky}},\ and\ \bibinfo
			{author} {\bibfnamefont {P.}~\bibnamefont {Zoller}},\ }\bibfield  {title}
		{\bibinfo {title} {{Strongly Correlated Gases of Rydberg-Dressed Atoms:
					Quantum and Classical Dynamics}},\ }\href
		{https://doi.org/10.1103/PhysRevLett.104.223002} {\bibfield  {journal}
			{\bibinfo  {journal} {Phys. Rev. Lett.}\ }\textbf {\bibinfo {volume} {104}},\
			\bibinfo {pages} {223002} (\bibinfo {year} {2010})}\BibitemShut {NoStop}%
		\bibitem [{\citenamefont {Johnson}\ and\ \citenamefont
			{Rolston}(2010)}]{Johnson2010rydberg}%
		\BibitemOpen
		\bibfield  {author} {\bibinfo {author} {\bibfnamefont {J.~E.}\ \bibnamefont
				{Johnson}}\ and\ \bibinfo {author} {\bibfnamefont {S.~L.}\ \bibnamefont
				{Rolston}},\ }\bibfield  {title} {\bibinfo {title} {{Interactions between
					Rydberg-dressed atoms}},\ }\href {https://doi.org/10.1103/PhysRevA.82.033412}
		{\bibfield  {journal} {\bibinfo  {journal} {Phys. Rev. A}\ }\textbf {\bibinfo
				{volume} {82}},\ \bibinfo {pages} {033412} (\bibinfo {year}
			{2010})}\BibitemShut {NoStop}%
		\bibitem [{\citenamefont {Min\'a\ifmmode~\check{r}\else \v{r}\fi{}}\ \emph
			{et~al.}(2022)\citenamefont {Min\'a\ifmmode~\check{r}\else \v{r}\fi{}},
			\citenamefont {van Voorden},\ and\ \citenamefont
			{Schoutens}}]{minar2022rydberg}%
		\BibitemOpen
		\bibfield  {author} {\bibinfo {author} {\bibfnamefont {J.~c.~v.}\
				\bibnamefont {Min\'a\ifmmode~\check{r}\else \v{r}\fi{}}}, \bibinfo {author}
			{\bibfnamefont {B.}~\bibnamefont {van Voorden}},\ and\ \bibinfo {author}
			{\bibfnamefont {K.}~\bibnamefont {Schoutens}},\ }\bibfield  {title} {\bibinfo
			{title} {Kink dynamics and quantum simulation of supersymmetric lattice
				hamiltonians},\ }\href {https://doi.org/10.1103/PhysRevLett.128.050504}
		{\bibfield  {journal} {\bibinfo  {journal} {Phys. Rev. Lett.}\ }\textbf
			{\bibinfo {volume} {128}},\ \bibinfo {pages} {050504} (\bibinfo {year}
			{2022})}\BibitemShut {NoStop}%
		\bibitem [{\citenamefont {B\"{u}chler}\ \emph {et~al.}(2007)\citenamefont
			{B\"{u}chler}, \citenamefont {Micheli},\ and\ \citenamefont
			{Zoller}}]{Bchler2007three}%
		\BibitemOpen
		\bibfield  {author} {\bibinfo {author} {\bibfnamefont {H.~P.}\ \bibnamefont
				{B\"{u}chler}}, \bibinfo {author} {\bibfnamefont {A.}~\bibnamefont
				{Micheli}},\ and\ \bibinfo {author} {\bibfnamefont {P.}~\bibnamefont
				{Zoller}},\ }\bibfield  {title} {\bibinfo {title} {Three-body interactions
				with cold polar~molecules},\ }\href {https://doi.org/10.1038/nphys678}
		{\bibfield  {journal} {\bibinfo  {journal} {Nature Physics}\ }\textbf
			{\bibinfo {volume} {3}},\ \bibinfo {pages} {726} (\bibinfo {year}
			{2007})}\BibitemShut {NoStop}%
		\bibitem [{\citenamefont {Gambetta}\ \emph {et~al.}(2020)\citenamefont
			{Gambetta}, \citenamefont {Li}, \citenamefont {Schmidt-Kaler},\ and\
			\citenamefont {Lesanovsky}}]{Gambetta2020engineering}%
		\BibitemOpen
		\bibfield  {author} {\bibinfo {author} {\bibfnamefont {F.~M.}\ \bibnamefont
				{Gambetta}}, \bibinfo {author} {\bibfnamefont {W.}~\bibnamefont {Li}},
			\bibinfo {author} {\bibfnamefont {F.}~\bibnamefont {Schmidt-Kaler}},\ and\
			\bibinfo {author} {\bibfnamefont {I.}~\bibnamefont {Lesanovsky}},\ }\bibfield
		{title} {\bibinfo {title} {{Engineering NonBinary Rydberg Interactions via
					Phonons in an Optical Lattice}},\ }\href
		{https://doi.org/10.1103/PhysRevLett.124.043402} {\bibfield  {journal}
			{\bibinfo  {journal} {Phys. Rev. Lett.}\ }\textbf {\bibinfo {volume} {124}},\
			\bibinfo {pages} {043402} (\bibinfo {year} {2020})}\BibitemShut {NoStop}%
		\bibitem [{\citenamefont {Myerson-Jain}\ \emph {et~al.}(2022)\citenamefont
			{Myerson-Jain}, \citenamefont {Yan}, \citenamefont {Weld},\ and\
			\citenamefont {Xu}}]{MyersonJain2022fractal}%
		\BibitemOpen
		\bibfield  {author} {\bibinfo {author} {\bibfnamefont {N.~E.}\ \bibnamefont
				{Myerson-Jain}}, \bibinfo {author} {\bibfnamefont {S.}~\bibnamefont {Yan}},
			\bibinfo {author} {\bibfnamefont {D.}~\bibnamefont {Weld}},\ and\ \bibinfo
			{author} {\bibfnamefont {C.}~\bibnamefont {Xu}},\ }\bibfield  {title}
		{\bibinfo {title} {{Construction of Fractal Order and Phase Transition with
					Rydberg Atoms}},\ }\href {https://doi.org/10.1103/PhysRevLett.128.017601}
		{\bibfield  {journal} {\bibinfo  {journal} {Phys. Rev. Lett.}\ }\textbf
			{\bibinfo {volume} {128}},\ \bibinfo {pages} {017601} (\bibinfo {year}
			{2022})}\BibitemShut {NoStop}%
	\end{thebibliography}
\end{document}